\documentclass[%
 reprint,
superscriptaddress,
 amsmath,amssymb,
 aps,
 pra,
]{revtex4-2}

\usepackage{graphicx}% Include figure files
\usepackage{dcolumn}% Align table columns on decimal point
\usepackage{bm}% bold math
\usepackage{bbm}
\usepackage{xcolor}
\usepackage{threeparttable}
\usepackage{multirow}
\usepackage{makecell}
\usepackage[varg]{txfonts} % aff合并

\usepackage[colorlinks=true,linkcolor=blue,citecolor=blue,urlcolor=blue]{hyperref}% add hypertext capabilities
\begin{document}

\preprint{APS/123-QED}

\title{Improving the trainability of VQE on NISQ computers for solving portfolio optimization using convex interpolation}% Force line breaks with \\
\author{Shengbin Wang}
\altaffiliation{These authors contributed equally to this work.} %共一
%\email{@quantumwang@126.com}
\affiliation{Origin Quantum Computing Company Limited, Hefei, 230026, China}
\affiliation{Laboratory of Quantum Information, University of Science and Technology of China, Hefei, 230026, China}
\author{Guihui Li}
\altaffiliation{These authors contributed equally to this work.}
%\email{@guihuilee@stu.ouc.edu.cn}
\affiliation{Department of Electronic Information and Artificial Intelligence, West Anhui University, Lu’an, 237012, China}
\author{Zhimin Wang}
%\email{@wangzhimin@ouc.edu.cn}
\affiliation{College of Physics and Optoelectronic Engineering, Ocean University of China, Qingdao, 266100, China}
\author{Zhaoyun Chen}
%\email{@chenzhaoyun@iai.ustc.edu.cn}
\affiliation{Institute of Artificial Intelligence, Hefei Comprehensive National Science Center, Hefei, 230088, China}
\author{Peng Wang}
%\email{@pwang0110@mail.ustc.edu.cn}
\affiliation{Laboratory of Quantum Information, University of Science and Technology of China, Hefei, 230026, China}
\author{Yongjian Gu}
%\email{@guyj@ouc.edu.cn}
\affiliation{College of Physics and Optoelectronic Engineering, Ocean University of China, Qingdao, 266100, China}
\author{Yu-Chun Wu}
\email{@wuyuchun@ustc.edu.cn}
\affiliation{Laboratory of Quantum Information, University of Science and Technology of China, Hefei, 230026, China}
\affiliation{Institute of Artificial Intelligence, Hefei Comprehensive National Science Center, Hefei, 230088, China}
\affiliation{Anhui Province Key Laboratory of Quantum Network, University of Science and Technology of China, Hefei, 230026, China}
\author{Guo-Ping Guo}
\email{@gpguo@ustc.edu.cn}
\affiliation{Origin Quantum Computing Company Limited, Hefei, 230026, China}
\affiliation{Laboratory of Quantum Information, University of Science and Technology of China, Hefei, 230026, China}
\affiliation{Institute of Artificial Intelligence, Hefei Comprehensive National Science Center, Hefei, 230088, China}
\affiliation{Anhui Province Key Laboratory of Quantum Network, University of Science and Technology of China, Hefei, 230026, China}

\date{\today}% It is always \today, today,

\begin{abstract}
Solving combinatorial optimization problems using variational quantum algorithms (VQAs) might be a promise application in the NISQ era. However, the limited trainability of VQAs could hinder their scalability to large problem sizes. In this paper, we improve the trainability of variational quantum eigensolver (VQE) by utilizing convex interpolation to solve portfolio optimization. Based on convex interpolation, the location of the ground state can be evaluated by learning the property of a small subset of basis states in the Hilbert space. This enlightens naturally the proposals of the strategies of close-to-solution initialization, regular cost function landscape, and recursive ansatz equilibrium partition. The successfully implementation of a $40$-qubit experiment using only $10$ superconducting qubits demonstrates the effectiveness of our proposals. Furthermore, the quantum inspiration has also spurred the development of a prototype greedy algorithm. Extensive numerical simulations indicate that the hybridization of VQE and greedy algorithms achieves a mutual complementarity, combining the advantages of both global and local optimization methods. Our proposals can be extended to improve the trainability for solving other large-scale combinatorial optimization problems that are widely used in real applications, paving the way to unleash quantum advantages of NISQ computers in the near future. 
%\begin{description}
%\end{description}
\end{abstract}

\maketitle

%\tableofcontents

\section{\label{sec:intro}INTRODUCTION}
In the noisy intermediate-scale quantum (NISQ) era \cite{Preskill2018}, the primary focus is to seek for the practical applications that can efficiently leverage the potential quantum advantages \cite{Bharti2022,Lau2022}. The proposal of variational quantum algorithms (VQAs) \cite{PeruzzoVQE2014,FarhiQAOA2014} has garnered extensive attention in the fields of quantum simulation, quantum machine learning and combinatorial optimization \cite{Bharti2022,Cerezo2021,Tilly2022,Cerezo2022challenges,Qu2024experimental}.
There are plenty of researches that demonstrate the possible quantum advantages in the small or intermediate problem sizes \cite{PeruzzoVQE2014,Kandala2017,otterbach2017unsupervised,Gard2020_symmetry-preserving,Harrigan2021quantum,Mohseni2022ising,BLEKOS20241}. However, it cannot be taken for granted that the trainability of VQAs can be maintained for large-scale problems. A barren plateau (BP) \cite{McClean2018barren}, given by an exponentially vanishing gradient scaling as $O(1/2^n)$ with $n$ being the number of qubits, leads to an exponential resource requirement, $O(2^{2n})$, in the optimization process to find a way downward, thus nullifying the potential quantum advantages. Additionally, a bumpy cost function landscape also hinders the trainability of local quantum models with shallow ansatze and local cost functions \cite{anschuetz2022quantum_pumpy_landscape}.

Recent progress for alleviating, and even possibly avoiding, the notorious BPs mainly focuses on the study of ansatze \cite{McClean2018barren,Nakaji2021BP_ALA_HEA,Wang2021noiseinduced,Tuysuz2023product_pqc}, parameter initializations \cite{McClean2018barren,Grant2019initialization,volkoff2021correlatedparameters,Holmescorrelationinitial}, cost functions \cite{CerezoBP2021,Amaro2022filtering,arrasmith2022equivalenceBPNG}, and optimization methods \cite{natural_gradient,Arrasmith2021effectofbarren,cerezo2021higherorder,Singh2023benchmarkingoptimizationmethod}. These strategies attempt to reduce the randomness of ansatze, the size of search spaces, and the accumulation of hardware noises in a more generic manner. Good parameter initializations can also alleviate the impact of the bumpy cost function landscape \cite{anschuetz2022quantum_pumpy_landscape}. For large problem sizes, it is apparent that the combination of multiple strategies should be the feasible way to activate the advantages of the current NISQ computers. In addition, problem-specific properties can provide some intrinsic insights (prior knowledge) for developing more efficient strategies to improve the trainability.

In this paper, we improve the trainability of variational quantum eigensolver (VQE) in the NISQ era by predicting the location of the optimal solution using convex interpolation \cite{Taylor2017convex_interpolation}, further advancing its practical application process for solving portfolio optimization \cite{MugelPRRpo,buonaiuto2023best,wang2024vqe-po} on NISQ computers. \textbf{Convex interpolation} is defined as follows: Given a set of data points $(x_0,y_0)$, $(x_1,y_1)$, $…$, $(x_{m-1},y_{m-1})$, where the values $y_i$, with $i \in [0,m-1]$, are arranged in a convex configuration, convex interpolation constructs an interpolated function $f(x)$ that passes through all the given data points $(x_i,y_i)$. The convexity of $f(x)$ can be expressed mathematically as 
$f(\lambda x_i +(1-\lambda) x_j) \leq \lambda f(x_i) +(1-\lambda)f(x_j)$ for any $\lambda \in [0,1]$ and integers $i,j \in [0,m-1]$.

The idea stems from the following observation that is closely associated with combinatorial optimization: A Dicke state \cite{wang2024vqe-po,Dicke1954Dicke_state,Bärtschi2022Dicke_state} possesses an inherent clustering property based on Hamming distance between its basis states. A generalized Dicke state $\vert D^n_k\rangle$ is a superposition state composed of $\binom{n}{k}$ $n$-qubit basis states $\vert i\rangle$ with Hamming weight $HW(i)=k$
\begin{equation}
\vert D^n_k\rangle = \sum\nolimits_{i \in \{0, 1\}^n,HW(i)=k} a_i\vert i\rangle,
\label{eq:Dicke state definition}
\end{equation}
where $a_i$ is the amplitude of basis state $\vert i\rangle$. This kind of Dicke state can be prepared by the CCC ansatz composed of multiple ``CNOT, C-RY, CNOT'' unitary blocks, with $k$ NOT gates on alternating qubtis as the input, proposed in our previous work \cite{wang2024vqe-po}. Please refer to Appendix \ref{appen:Principle analysis of the observation}.1 for the theoretical insight of the clustering property of the ansatz. In Appendices \ref{appen:Proof product decomposition of CCC ansatz} and \ref{appen: completeness and reachability proves}, the observation is proved rigorously that a CCC ansatz can be partitioned into multiple sub-ansatze based on the Hamming distance between the basis states in different sub-ansatze. In the proof, the Hamming distance is expressed by the Hamming weight of each product part of the sub-ansatz. It implicitly indicates the restricted expressibility of the shallow CCC ansatz, which could improve the trainability of VQAs on NISQ hardwares. Therefore, on the global perspective of the cost function landscape, the energy of a basis state $\vert b\rangle$ with a larger Hamming distance from the ground state $\vert b_g\rangle$ results in a greater energy gap away from the ground state energy, i.e. $\Delta E(\vert b\rangle, \vert b_g\rangle)\propto HD(\vert b\rangle, \vert b_g\rangle)$, in the overall distribution trend. Specifically, the Hamming distance indicates the difference between a given asset allocation and the optimal asset allocation. The greater the difference, the higher the energy of the corresponding asset allocation. Inspired by the observation, the distribution of the minimal energies of the sub-ansatze should intuitively obey a convex distribution. 

In the present scenario, the data points $(x_i,y_i)$ correspond to the optimal states $\vert b_i \rangle$ and their respective minimum energies $E(\vert b_i \rangle)$ for the $m$ subspaces: $(\vert b_0 \rangle, E(\vert b_0 \rangle))$, $(\vert b_1 \rangle, E(\vert b_1 \rangle))$, $…$, $(\vert b_g \rangle, E(\vert b_g \rangle))$, …, $(\vert b_{m-1} \rangle, E(\vert b_{m-1} \rangle))$. Our objective is not to use these points to interpolate a convex function but rather to utilize a small subset of easily obtainable points to estimate the location of the function's minimum. Here, the term ``location'' refers to the subspace region where $(\vert b_g \rangle, E(\vert b_g \rangle))$ resides, rather than the exact solution itself. Furthermore, our intuition suggests that convexity may reduce to monotonicity. Monotonicity here refers to the relationship between the minimum energies of the subspaces, where $E(\vert b_i \rangle) > E(\vert b_j \rangle)$ (or $E(\vert b_i \rangle) < E(\vert b_j \rangle)$) when $i<j \leq g$ (or $g \leq i<j$). In consequence, interpolating a convex curve using the minimal energies of a small number of subspaces (sub-ansatze) would allow us to estimate the location of the ground state. In Appendix \ref{appen:Principle analysis of the observation}.2, we thoroughly analyzed the strong validity of our intuition both theoretically and numerically. Even so, our intuition can be regarded as an approximate convex interpolation in the small number of extreme cases, through which the location of the ground state subspace can still be efficiently predicted.

The ability to predict the ground state location indicates that the parameters $\bm{\theta}$ can be intentionally initialized close to the solution $\vert b_g\rangle$. Hence, the fidelity $F=|\langle\psi(\bm{\theta})\vert b_g\rangle|^2$ can be lower bounded. From the cost function point of view, the shape of the cost function landscape is also determined by the relationship between assets, $\bm{R}$, as formulated within the cost function $C(\bm{\theta}, \bm{R})$. A reasonable adjustment of this relationship should be able to produce a much smoother and more regular landscape. This leads to the proposal of efficient strategies for close-to-solution initialization and regular cost function landscape, aiming to reduce the randomness of the ansatz, decrease the size of the search space, and enhance the capability of escaping from the spurious local minima. Moreover, a Dicke state can be decomposed into smaller separable Dicke states recursively. Hence, the location of the ground state can be evaluated using a smaller quantum computer. This motivates the development of the recursive ansatz equilibrium partition method. The size of the search space is recursively reduced, and the impact of hardware noise is drastically alleviated. By combining the three proposed advances, we successfully obtained the optimal solution for a $40$-qubit experiment using only $10$ qubits on our superconducting quantum computer ``Wukong'' \cite{Dou2022QPanda}. While the convex interpolation idea primarily originates from the quantum perspective, it also inspired the development of a prototype (classical) greedy algorithm.

As can be seen, our contribution to improving trainability in the NISQ era lies in a complete scheme that integrates ansatz design, parameter initialization, and cost-function selection, which were previously investigated only separately. Our proposals can be applied to solve other large-scale combinatorial optimization problems, such as maxcut, number partitioning, markit split \cite{Lucas2014ising,NanniciniPRE}, and the more general graph partition problems \cite{graph_partition}. Particularly, the crucial initial step in distributed quantum computing \cite{Caleffi2022dqc} is to partition a monolithic quantum algorithm into different quantum computers for execution. The recursive ansatz partition method, using several small-scale quantum computers, can partition a large-scale quantum algorithm and simultaneously enable it to be solved on these small-scale quantum computers. This eliminates the dependence of algorithm partitioning on large-scale classical algorithms or large-scale quantum computers.

The structure of the paper is as follows. Firstly, a brief review of the portfolio optimization problem is provided in Section \ref{sec:port op}. Then, the details of the convex interpolation idea and its inspired close-to-solution initialization, regular cost function landscape and recursive ansatz equilibrium partition strategies are discussed in Section \ref{sec:methods}. Afterwards, Section \ref{sec:Experiments} presents the configurations and methods developed for the successful executions of our experiments for solving portfolio optimization and graph bisection, including the $40$-qubit portfolio optimization experiment. Finally, we conclude our work and discuss some interesting problems in Section \ref{sec:Conclusion and discussion}.

\section{\label{sec:port op}Portfolio optimization overview}
Portfolio optimization is one of the most common optimization problems in finance \cite{MugelPRRpo,buonaiuto2023best,wang2024vqe-po,Orús2019_finance,Herman2022_finance}. Its objective is to achieve the expected return with minimum financial risk, or to maximize the expected return for a given level of risk, by selecting from an asset pool a set of optimally allocated assets. The binary integer portfolio optimization is modeled as
\begin{equation}
\begin{split}
    \operatorname{min}&_{\bm{x}}\ {q\bm{x}^TA\bm{x}}-\bm{\mu}^T\bm{x},\\
    s.t.& \ \xi =\bm{\Pi}^T\bm{x}, \label{eq:mean-variance}
\end{split}
\end{equation}
where $\bm{x}=(x_1,x_2,\ldots,x_n)^T$, with $x_i\in\{0, 1\}$, represents the asset selection vector, and $n$ denotes the number of assets in the asset pool. The \emph{i}th asset is selected when the binary variable $x_i=1$ or not selected when $x_i=0$. $A$ is the real covariance matrix, $q>0$ is the risk level, and the vector $\bm{\mu}$ contains the expected returns of the assets. The relationship between assets, $\bm{R}$, is stored in $q$, $A$, and $\bm{\mu}$. In the constraint term, $\xi$ is the investor's budget, corresponding to $k$ in Dicke state $\vert D^n_k\rangle$. And vector $\bm{\Pi}$ encompasses the prices of the assets. In this paper, vector $\bm{\Pi}$ is simplified to the all-ones vector $\mathbbm{1}$ of dimension $n$.

The problem is typically mapped into a quadratic unconstrained binary optimization (QUBO) problem for solution. In our previous work, the constraint term in Eq. \eqref{eq:mean-variance} is regarded as a hard constraint and is encoded into the construction of CCC ansatz \cite{wang2024vqe-po}. From the functionality point of view, the CCC ansatz is a Dicke state preparation ansatz. The asset selection vector $\bm{x}$ becomes a trigonometric function of parameters $\bm{\theta}$. As a result, the cost function degenerates to the minimization term of Eq. \eqref{eq:mean-variance} only, $C(\bm{\theta}, \bm{R})$. It naturally corresponds to an Ising model with external magnetic field. Using the substitution $x_i = \frac{1 - z_i}{2}$, binary variables $x_i\in \{0,1\}$ can be converted to spin variables $z_i\in\{+1,-1\}$. The problem can be modeled as
\begin{equation}
\operatorname{min}_{\bm{z}}\ {q^{'}\bm{z}^TA\bm{z}}-\bm{\mu}^{{'}T}\bm{z}, \label{eq:Ising hamiltonian}
\end{equation}
where $\bm{z} = (z_1,z_2,\ldots,z_n)^T$ is the asset selection vector in the spin variable form. The new risk level $q^{'}$ equals $\frac{q}{4}$, and the new expected return vector $\bm{\mu}^{'}$ equals $\frac{1}{2}(qA\mathbbm{1}-\bm{\mu})$, where $\mathbbm{1}$ is also the all-ones vector of dimension $n$. The constant factor $(\frac{q}{4}\mathbbm{1}^TA-\frac{1}{2}\bm{\mu}^T)\mathbbm{1}$ is omitted. The model is converted to the problem Hamiltonian by replacing the spin variable $z_i$ with the Pauli $Z$ operator operating on the $i$th qubit, denoted as $\sigma^Z_i$. Since the Hamiltonian is not encoded into the ansatz in the VQE, in the following sections, we present our results in the more natural binary variable form for simplicity and clarity.

\section{\label{sec:methods}Convex interpolation}
We found that by clustering based on the Hamming distance between the basis states, the Dicke state $\vert D^n_k\rangle$ can be decomposed into a linear combination of $({\rm min}\{j,k\}+1)$ product terms with certain amplitudes $a_i$ as
\begin{equation}
\begin{split}
\vert D^n_k\rangle = \sum^{\operatorname{min}\{j,k\}}_{i=0}a_i\vert D^j_i\rangle\otimes\vert D^{n-j}_{k-i}\rangle. \label{eq:equation 1}
\end{split}
\end{equation} %%%%%%%%%%%%%
Each product term is composed of two complementary Dicke states $\vert D^j_i\rangle$ and $\vert D^{n-j}_{k-i}\rangle$, where $j \leq n-j$ is the number of qubits that is arranged into the first part, $\vert D^j_i\rangle$, of each product term, refer to Appendix \ref{appen:Proof product decomposition of CCC ansatz} for the rigorous analysis. In the following, $\vert D^j_i\rangle$ and $\vert D^{n-j}_{k-i}\rangle$ are referred to as ``fragments'' of the product term. This provides us with a method that requires exponentially fewer samplings than the circuit cutting technique \cite{Peng2020_circuit_cutting, Harada2023optimalcircuitcuttingwithout} for solving large portfolio optimization problems using small quantum computers.

\begin{figure*}[ht]
    \centering    \includegraphics[width=1.0\textwidth]{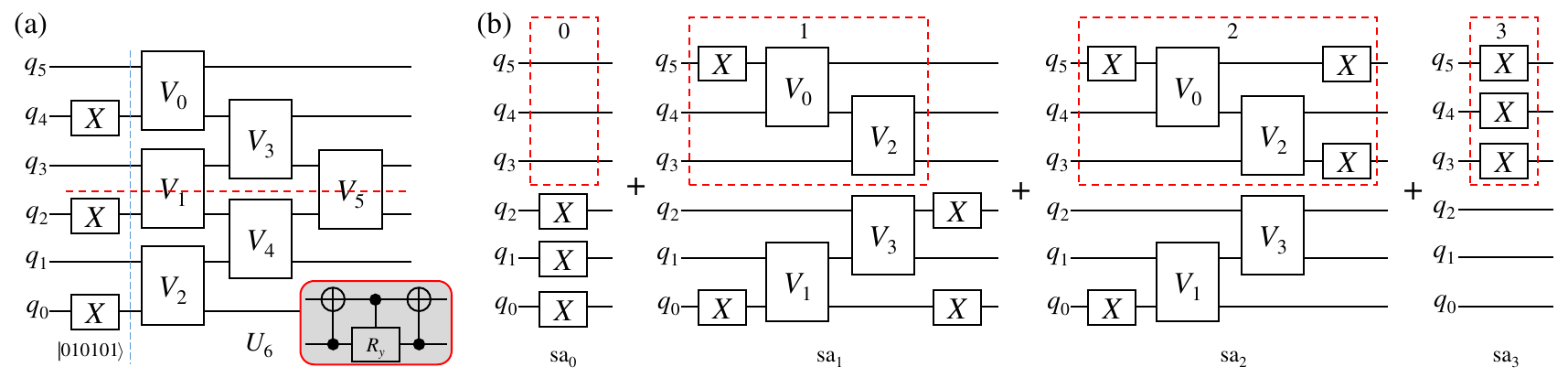}
    \caption{(a) The CCC ansatz for preparing the Dicke state $\vert D^6_3\rangle$. Each $V_i$ is constructed using a ``CNOT, C-RY, CNOT'' ($3$C) block depicted in the red box. The parameter is the angle of C-RY gate. Column $010101$ of the unitary $U_6$ contains all $\binom{6}{3}$ basis states. (b) The direct product decomposition of the CCC ansatz in (a). The entanglement between the two fragments are fully decomposed into $4$ product sub-ansatze. The red dashed boxes indicate the increase of $X$ gates in the upper fragment. In the lower fragment of sa$_1$ and the upper fragment of sa$_2$, the Dicke state $\vert D^3_2\rangle$ is prepared based on the relationship $\vert D^n_k\rangle = X^{\otimes n}\vert D^n_{n-k}\rangle$, as detailed in Appendix \ref{appen: completeness and reachability proves}.}
    \label{fig:figure prodecomp}
\end{figure*}

In this paper, we mainly focus on discussing the Dicke state $\vert D^n_{n/2}\rangle$ with an even $n$. The number of basis states it contains, $\binom{n}{n/2}$, is approaching $2^n$, making it the most complex $n$-qubit Dicke state. The results can be readily generalized to other relatively simple cases. The folded staircase structure of the CCC ansatz for preparing the Dicke state $\vert D^n_{n/2}\rangle$ is illustrated in Fig. \ref{fig:figure prodecomp}(a) by the circuit used for preparing $\vert D^6_3\rangle$. Refer to Appendix \ref{appen: completeness and reachability proves} for the rigorous proof of the completeness and reachability of CCC ansatze. Where $q_{n-1}$ is the highest qubit and $q_{0}$ is the lowest qubit. The initial state, $\vert 0101...01\rangle$, is prepared by the $X$ gates in the first time step. In this paper, we refer to the ``1'' elements in $\vert 0101...01\rangle$ as the ``$1$'' information, which indicates the selected assets or qubits. Subsequently, the ``$1$'' information in the initial state spreads throughout the entire search space via the unitary $U_n$, constructed by the $(\frac{n^2}{8}+\frac{n}{4})$ parameterized $2$-qubit unitaries $V_i$.

\begin{figure*}[ht]
    \centering
    \includegraphics[width=0.75\textwidth]{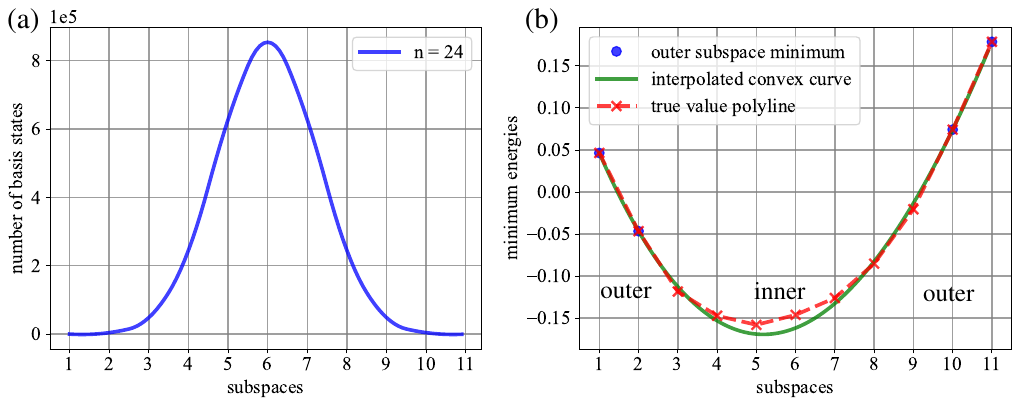}
    \caption{(a) The number of basis states distributed in each sub-ansatz for $\vert D^{24}_{12}\rangle$. It resembles a Gaussian distribution. The sub-ansatze closing to the edge, i.e. sa$_1$ and sa$_{11}$, contain a much smaller number of basis states. (b) The illustration of the convex interpolation. The red dashed polyline connects the true minimum energies of the sub-ansatze, represented by the red \textcolor{red}{×} marks. The green solid line represents the interpolated convex curve using the minimum energies of the $4$ outer sub-ansatze, denoted by the blue dots, using $scipy.interpolate.KroghInterpolator()$.}
    \label{fig:figure dis&curve}
\end{figure*}

In the most complex case, Eq. \eqref{eq:equation 1} can be reformulated as
\begin{equation}
\begin{split}
\vert D^n_{n/2}\rangle = \sum^{n/2}_{i=0} a_i \vert D^{n/2}_{i}\rangle \otimes \vert D^{n/2}_{n/2-i}\rangle, \label{eq:Dicke linear combination}
\end{split}
\end{equation}
where each $n$-qubit product state is composed of two $\frac{n}{2}$-qubit Dicke states. In this way, the CCC ansatz can be partitioned into $(\frac{n}{2}+1)$ sub-ansatze, as illustrated in Fig. \ref{fig:figure prodecomp}(b). Each $n$-qubit sub-ansatz is a direct product of two $\frac{n}{2}$-qubit fragments. That is to say, the search space $H$, spanned by the CCC ansatz, is partitioned into $(\frac{n}{2}+1)$ product subspaces as $H=\sum ^{n/2}_{i=0} (\bigotimes^1_{j=0} H_{i,j})$. In other words, it is partitioned into $n+2$ fragments with $\frac{n}{2}$ qubits. The $\frac{n}{2}$-qubit $H_{i,j}$ enables the simulation of the $n$-qubit sub-ansatz by an $\frac{n}{2}$-qubit quantum computer directly. Comparatively, the circuit cutting technique  \cite{Harada2023optimalcircuitcuttingwithout} would introduce $4^{\frac{n}{2}}$ (or $4^{\frac{n}{2}+1}$) fragments, as cutting an $n$-qubit CCC ansatz into multiple $\frac{n}{2}$-qubit fragments requires cutting $\frac{n}{2}$ (or $\frac{n}{2}+1$) CNOT gates, when $\frac{n}{2}$ is even (or odd), as illustrated in Fig. \ref{fig:figure prodecomp}(a). The $(\frac{n}{2}+1)$ sub-ansatze, sa$_0$, sa$_1$, …, and sa$_{\frac{n}{2}}$, correspond to the distributions $(0, \frac{n}{2})$, $(1, \frac{n}{2}-1)$, …, and $(\frac{n}{2}, 0)$, where $i$ and $\frac{n}{2}-i$ in $(i, \frac{n}{2}-i)$ represent the number of $X$ gates distributed to the upper and lower fragments, respectively. In other words, the subscript of a sub-ansatz indicates the number of $X$ gates distributed to its upper fragment. Extensive numerical simulations on both real and random datasets demonstrate that the minimum energies of these sub-ansatze can be used to construct a convex polyline, or one that is very nearly so (see Fig. \ref{fig:figure dis&curve}(b) for a simple illustration).

The above results inspired the proposal of the convex interpolation idea. Specifically, using the convex interpolation idea to evaluate the location of the ground state requires the minimum energies of a small number of subspaces. Clearly, a fragment with either a dense or sparse distribution of $X$ gates results in a sparse superposition state, i.e. $\binom{n/2}{i}=\binom{n/2}{n/2-i}$. In consequence, as illustrated in Fig. \ref{fig:figure dis&curve}, the minimum energies of the outer sub-ansatze, sa$_1$, sa$_2$, sa$_{\frac{n}{2}-2}$, and sa$_{\frac{n}{2}-1}$, are the simplest and most effective choices due to the smaller number of basis states they contain. Sub-ansatze sa$_0$ and sa$_{\frac{n}{2}}$ are ignored because they only contain one basis state. In the present proof of principle scenario, the brute force method is utilized to acquire the minimum energies of the four outer sub-ansatze. In addition, we also revise the interpolated convex curve using empirical information. For states $\vert D^{n/2}_{i}\rangle$ and $\vert D^{n/2}_{n/2-i}\rangle$, which contain the same number of basis states, the total number of basis states of the $4$ outer sub-ansatze is $2(\binom{n/2}{1}^2+\binom{n/2}{2}^2)$, resulting in an asymptotic complexity of $O(n^4)$ with a constant coefficient of $\frac{1}{32}$. For instance, interpolating $\vert D^{100}_{50}\rangle$ requires $3\times10^6$ basis states. Hence, for larger problem sizes, the proposed recursive partition method would be more effective, refer to subsection \ref{subsec:recursive ansatz equilibrium partition} for details.

\begin{figure}[ht]
    \centering
    \includegraphics[width=0.35\textwidth]{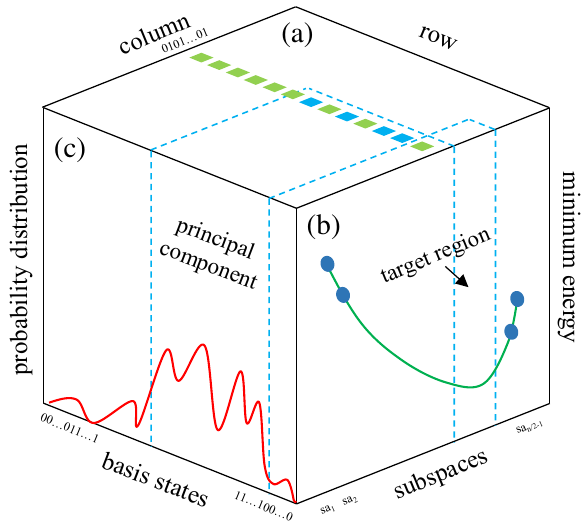}
    \caption{(a) The matrix representation of column $0101...01$ of the unitary $U_n$, with colored squares representing its elements. (b) The interpolated convex curve. The target region contains the ground state. The subspaces within the target region (between the blue dashed line) encompass the basis states corresponding to the blue squares in (a). (c) The initialized probability distribution. Its principal component (between the blue dashed lines) covers the target region.}
    \label{fig:figure relationmatrix}
\end{figure}

After evaluating the location of the ground state using the interpolated convex curve, the CCC ansatz, see Fig. \ref{fig:figure prodecomp}(a), can be initialized to produce a probability distribution with its principal component covering the location. As shown in Fig. \ref{fig:figure relationmatrix}, the relationship between the principal component of the probability distribution and the interpolated ground state location (the target region) is connected by column $0101...01$. This is referred to as the close-to-solution initialization strategy. This strategy can efficiently improve the trainability of the VQAs. Now the question is how to find the parameter configurations that can consistently produce various probability distributions whose principal component can regularly cover different locations of the search space.

However, in large-size problems, the interpolated convex curve may not always perfectly match the true value polyline. The main reason is that the cost function landscape, determined by the random relationships between the assets, is not smooth enough. In some cases, this results in a grotesque convex polyline of the true value which cannot be interpolated effectively by merely improving the interpolation method. Hence, the non-smoothness of the cost function landscape is an essential issue limiting the trainability of VQAs. However, the shallow ansatz inherently impedes the smoothing of the cost function landscape \cite{anschuetz2022quantum_pumpy_landscape}. Drawing inspiration from the convex interpolation idea, the next best thing is to achieve a more regular distribution of the local minima. Therefore, the second question is how to adjust the relations between the assets to obtaining a regular cost function landscape.

Furthermore, each fragment of the product term $\vert D^j_i\rangle\otimes\vert D^{n-j}_{k-i}\rangle$ in Eq. \eqref{eq:equation 1} is also a Dicke state. It can be decomposed into a linear combination of the product of several smaller Dicke states in a recursive manner, until there are no degrees of freedom left for further decomposition. Consequently, the ground state can be evaluated using the sub-ansatz predicted to contain the ground state, rather than the original CCC ansatz. The recursive ansatz partition method reduces the size of the search space, and releases the hardware constraints on qubit number and circuit depth, improving the trainability of VQE on NISQ computers. However, in order to obtain potential quantum advantages, the suitable number of partitions should be analyzed critically.

Convex interpolation idea also inspired the development of a prototype greedy algorithm, elucidating the rationale behind the effectiveness of the greedy algorithm. Based on the Hamming distance-based clustering property, the search subspace is expected to gradually transfer to the ground state location. That is to say, the greedy algorithm has the ability to transfer among the partitioned subspaces. However, the greedy algorithm mainly focuses on the minimum of the current subspace. The transfer may become stuck when the landscape of the current local minimum is too steep. This reflects its locality. Multiple initial basis states or a more regular cost function can alleviate the issue.

In the following three subsections, we introduce the close-to-solution initialization, regular cost function landscape, and recursive ansatz equilibrium partition strategies, which are designed to improve the trainability of VQE for solving portfolio optimization.

\subsection{\label{subsec:close-to-solution initialization}Close-to-solution initialization}
The shallow CCC ansatz has constrained expressibility, constructing a bumpy cost function landscape. Yet, it encompasses all $\binom{n}{n/2}$ basis states, ensuring basis states completeness. This implies that the dimension of the search space is exponential to $n$, restricting the effective trainability of the VQE. Leveraging the convex interpolation idea, efficient trainability can be achieved if the principal component of the probability distribution prepared by the ansatz precisely covers the ground state location. However, the probability distribution is also bumpy, which hinders rigorous derivation.

We find that the relationship between the parameters and the principal component of the Dicke state depends on how the propagation of the ``$1$'' information occurs through the folded staircase CCC ansatz, refer to Appendix \ref{appen:principal component upon parameters}. The main results can be summarized as two phases: (1) Global principal component distribution. The first layer of $V_i$s applied in the second time step, see Fig. \ref{fig:figure prodecomp}(a), achieves the propagation of $1$s on a global scale and, simultaneously, obtains the principal component distribution form of the Dicke state upon identical parameter $\theta \in[0,\pi]$; (2) Local completeness propagation. The subsequent $V_i$s in the following time steps enrich the completeness of the basis states of the Dicke state layer by layer through contractive local propagation. The two phases together achieve the unification of the principal component distribution and the completeness of basis states. The situation is similar for Dicke state $\vert D^n_k\rangle$ with $k<n/2$.

\begin{figure*}[ht]
    \centering
    \includegraphics[width=0.9\textwidth]{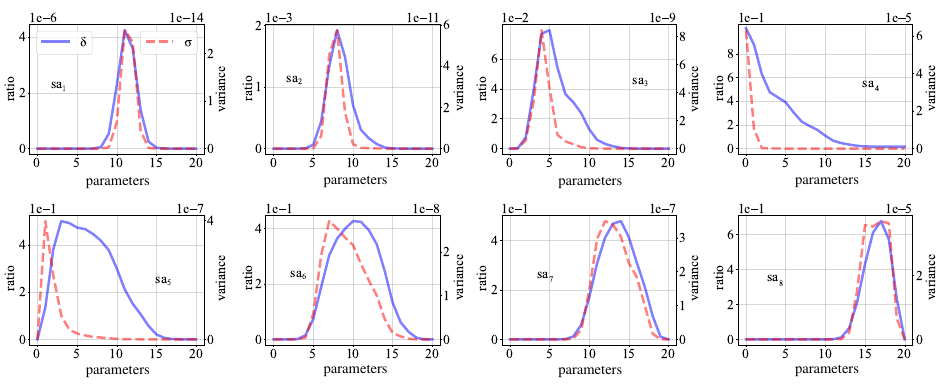}
    \caption{The ratio and variance curves of sub-ansatze sa$_1$ to sa$_8$ of $\vert D^{18}_9\rangle$ with respect to the $21$ parameter values sampled uniformly on the interval $[0,\pi]$, i.e. $0$, $\pi/20$, $\pi/10$, $...$, $\pi$. Each sub-ansatz (principal component) has its own optimal ratio interval, i.e. the interval producing the larger ratios, corresponding to the concave (hill) part of the curves. As the subscript of the sub-ansatz increases from $1$ to $\frac{n}{2}-1$, the optimal ratio interval migrates to the small parameter and then to the large parameter. This is consistent with the propagation mode of the ``$1$'' information. The variance changes in basically the same way as the ratio. The values of the variances may seem small, e.g. $10^{-5}$, yet they are compressed by BP. This is in accordance with the exponential vanishing $1/2^{18}\approx 10^{-5}$. The initial state $\bigotimes^9_{i=1}\vert 01\rangle$ is assigned to sa$_4$. This results in a ratio close to $1$ when the parameter is initialized not far from $0$. However, in this scenario, the principal component is mainly focused on basis state $\bigotimes^9_{i=1}\vert 01\rangle$.}
    \label{fig:figure close-to-solution initialization}
\end{figure*}

We introduce the ratio, $\delta$, and variance, $\sigma$, of the principal component as metrics to assess the performance (concentration and smoothness of the principal component) of the close-to-solution initialization strategy. Here the principal component refers to the set of probabilities corresponding to the basis states produced by each sub-ansatz. For sub-ansatz sa$_i$, the ratio $\delta_i$ is
\begin{equation}
\begin{split}
\delta_i = \sum_{\vert b_j\rangle\in \vert {\rm sa}_i\rangle} p_{\vert b_j\rangle} / \sum_{\vert b_h\rangle\in\vert D^n_{n/2}\rangle}p_{\vert b_h\rangle} = \sum_{\vert b_j\rangle\in \vert {\rm sa}_i\rangle} p_{\vert b_j\rangle}, 
\label{eq:ratio}
\end{split}
\end{equation}
where $\vert {\rm sa}_i\rangle = \vert D^{n/2}_{i}\rangle\otimes \vert D^{n/2}_{n/2-i}\rangle$ represents the state prepared by sub-ansatz sa$_i$, and $\vert b_j\rangle$ and $\vert b_h\rangle$ represent basis states in $\vert {\rm sa}_i\rangle$ and $\vert D^n_{n/2}\rangle$, respectively. In other words, $\delta_i$ is the probability of measuring the basis states of the principal component $\vert {\rm sa}_i\rangle$. The variance $\sigma_i$ is
\begin{equation}
\begin{split}
\sigma_i = \frac{1}{N}\sum_{\vert b_j\rangle\in \vert {\rm sa}_i\rangle}(\frac{\delta_i}{N} - p_{\vert b_j\rangle})^2,
\label{eq:variance}
\end{split}
\end{equation}
where $N=\binom{n/2}{i}\binom{n/2}{n/2-i}$ is the total number of basis states produced by sa$_i$.

Intuitively, larger ratio and smaller variance are expected. However, it is regrettable that, for a shallow ansatz, such expectation is paradoxical. As depicted in Fig. \ref{fig:figure close-to-solution initialization}, the changes of the two metrics are essentially synchronous. This arises from the bumpy nature of the probability distribution, which severely constrains the smoothness of the principal component. A small variance is usually associated with a small ratio, necessitating a trade-off between smoothness and concentration. Conversely, a high ratio implies a higher concentration of the principal component on fewer basis states, resulting in a larger variance. In consequence, achieving an effective compromise between ratio and variance requires selecting parameters corresponding to the hillside of the concave part of the curves.  However, sub-ansatze with excessively small ratios and variances that do not form a principal component, such as sa$_1$ to sa$_3$, should also be rejected.

\begin{figure}[ht]
    \centering
    \includegraphics[width=0.4\textwidth]{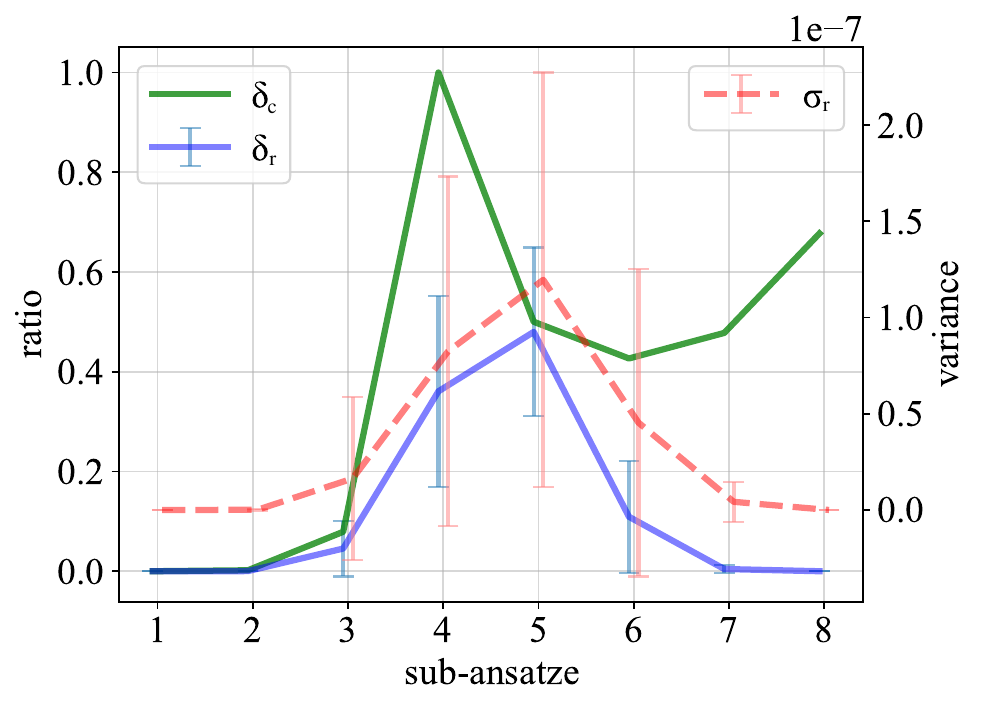}
    \caption{Comparison between close-to-solution initialization and random initialization. $\delta_c$ represents the optimal ratio of each sub-ansatz based on close-to-solution initialization, while $\delta_r$ (with error bars) represents the average ratio, and $\sigma_r$ (with error bars) represents the average variance of the $20$ simulations based on random initialization. As can be seen, close-to-solution initialization is overall better than random initialization.}
    \label{fig:figure comp2random}
\end{figure}

Here, we compare the proposed close-to-solution initialization with random initialization by randomly initializing the parameters $20$ times on the interval $[0,\pi]$, as shown in Fig. \ref{fig:figure comp2random}. As depicted in the figure, for the random initialization scenario, only sub-ansatze sa$_4$ and sa$_5$ in the middlemost region have sufficient large ratios. Even so, their standard derivations are large, introducing a significant amount of uncertainty. Intriguingly, for the right half of the sub-ansatze, from sa$_{\frac{n}{4}}$ to sa$_{\frac{n}{2}-1}$, the close-to-solution initialization is far superior to random initialization, especially for sub-ansatze close to sa$_{\frac{n}{2}}$.

To sum up, the interpolated ground state should be located within the right half of the sub-ansatze. This can be achieved by reversing the order of the assets when the ground state is interpolated to be located in the first half of the sub-ansatze, from sa$_1$ to sa$_{\frac{n}{4}}$. Moreover, identical parameter initialization is compatible with the correlated parameter optimization strategy \cite{volkoff2021correlatedparameters,Holmescorrelationinitial}, providing an efficient method to alleviate the impact of the bumpy cost function landscape (local minima) and hardware noise. In fact, the regular cost function landscape strategy can easily interpolate the ground state to the right, naturally improving the efficiency of our close-to-solution initialization strategy. Therefore, the first question can be answered affirmatively.

\subsection{\label{subsec:regular cost function landscape}Regular cost function landscape}
Intuitively, the asset information stored in the risk level $q$, the covariance matrix $A$, and the expected return vector $\bm{\mu}$ in Eq. \eqref{eq:mean-variance} determine the shape of the cost function landscape. Regular rearrangement of this information should be capable of reshaping the bumpy cost function landscape into a more regular one. As a consequence, we could obtain a smaller number of local minima with gentler steepness, facilitating easier convergence to the ground state. In fact, graph partition problem can be solved by reordering the Laplacian matrix of the problem graph into its similarity matrix with a block diagonal form \cite{Mahmoud_reordering}. However, the total number of permutations is $n$! with $n$ the number of nodes in the graph, making it NP-hard.

\begin{figure*}[ht]
    \centering
    \includegraphics[width=0.9\textwidth]{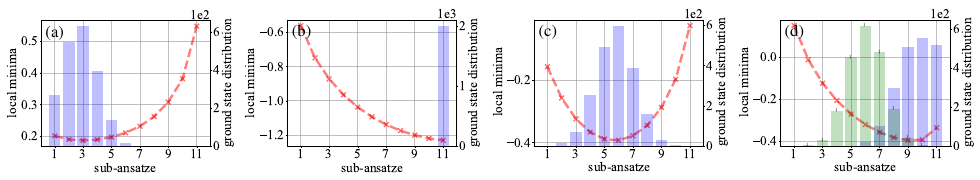}
    \caption{Average true value polylines (red dotted curves) and ground state distributions (blue bars) of $\vert D^{24}_{12}\rangle$, reordering based on (a) covariance matrix with cost function $\min_{\bm{\theta}} C(\bm{\theta}, A)$, (b) expected return vector with cost function $\min_{\bm{\theta}} C(\bm{\theta}, \bm{\mu})$, (c) covariance matrix with $\min_{\bm{\theta}} C(\bm{\theta}, q, A, \bm{\mu})$, and (d) expected return vector with $\min_{\bm{\theta}} C(\bm{\theta}, q, A, \bm{\mu})$. The constant factor of $10$ in the cost function is disregarded. Each polyline is plotted based on $2000$ simulations on both real and random datasets ($1000$ times each. Two types of datasets yield highly consistent results). The green bars with a black line on top in (d) represent the ground state distributions of random initialization on the interval $[0,\pi]$, similar to a Gaussian distribution. When reordered based on $\bm{\mu}$, $99.65\%$ of the ground states of the $2000$ simulations are biased to the right, whereas without reordering, only $35.8\%$ are biased to the right.}
    \label{fig:figure polyline&distri}
\end{figure*}

In portfolio optimization, the impact of the expected return vector $\bm{\mu}$ and risk level $q$ also need to be considered. For simplicity, in this paper, assets are reordered based on their variances, represented in the diagonal of the covariance matrix $A$, and by their expected returns in vector $\bm{\mu}$. In both cases, the reordering complexity is $O(n^2)$. In accordance with the actual situation, the expected return vector $\bm{\mu}$ is the principal component in the cost function commonly. Extensive numerical simulations confirm that expected return-based reordering leads to a more regular and biased cost function landscape compared to variance.

This can be analyzed in $4$ situations, as depicted in Fig. \ref{fig:figure polyline&distri}. The cost function $C(\bm{\theta}, \bm{R})$ with 
$\bm{R}$ formulated by different asset information, $A$, $\bm{\mu}$, or $(q, A, \bm{\mu})$, turns to
\begin{equation}
\begin{split}
C(\bm{\theta}, A) &= \bm{x}^TA\bm{x}, \\
C(\bm{\theta}, \bm{\mu}) &= -\bm{\mu}^T\bm{x}, \\
C(\bm{\theta}, q, A, \bm{\mu}) &= q\bm{x}^TA\bm{x} -\bm{\mu}^T\bm{x}.
\label{eq:cost function with diff R}
\end{split}
\end{equation}
The goal is to minimize $C(\bm{\theta}, A)$ and $C(\bm{\theta}, \bm{\mu})$ in the first two extreme situations. As shown in Fig. \ref{fig:figure polyline&distri}(a) and Fig. \ref{fig:figure polyline&distri}(b), the expected return $\bm{\mu}$ constitutes a full principal component compared to the variance (sa$_0$ and sa$_{12}$ are ignored). The ground state of $C(\bm{\theta}, \bm{\mu})$ is located in the rightmost subspace. Moreover, the minima produced by the covariance matrix $A$ are always larger than those produced by the expected return vector $\bm{\mu}$ in all $2000$ simulations on both real and random datasets. That is to say, the expected return vector $\bm{\mu}$ should play a dominate role in $C(\bm{\theta}, q, A, \bm{\mu})$ when $q$ is upper bounded. As shown in Fig. \ref{fig:figure polyline&distri}(c) and Fig. \ref{fig:figure polyline&distri}(d), for the compounded cost function $C(\bm{\theta}, q, A, \bm{\mu})$ with $q = 1.0$, a neutralizing effect between the variances and expected returns emerges.

As the value of $q$ increases from $0$ to $\infty$, the cost function changes from $C(\bm{\theta}, \bm{\mu})$, through $C(\bm{\theta}, q, A, \bm{\mu})$, to $C(\bm{\theta}, A)$. The ground state location would gradually migrate to the middle (random distribution) starting from the rightmost subspace (Fig. \ref{fig:figure polyline&distri}(b)). Nonetheless, the ground state energy also increases synchronously. To produce a positively expected return, the ground state energy should be negative, i.e., the risk level $q$ is truly upper-bounded. This commonly leads to an interpolated curve with its minimum biased towards the subspaces with large subscripts on the right side. The expected return vector $\bm{\mu}$ constitutes the principal component of the compounded cost function $C(\bm{\theta}, q, A, \bm{\mu})$ in practice. This suggests that the second question can indeed be answered.

In summary, the expected return-based reordering is applied in our algorithm to achieve a more regular cost function landscape. The ground state is biased towards the sub-ansatze near the right interpolating points, resulting in a smaller search subspace and a more accurate interpolation. In consequence, the identical parameter $\theta$ can be accurately initialized and possess strong error resistance capability. Thus, the inherent compatibility of close-to-solution initialization and regular cost function landscape strategies greatly improves the trainability of VQE for solving portfolio optimization.

\subsection{\label{subsec:recursive ansatz equilibrium partition}Recursive ansatz equilibrium partition}
The close-to-solution initialization strategy distributes the principal component of the probability distribution to the target region by setting the parameters to specific values. This method can be considered a form of soft constraint approach, as it demands only a low interpolation accuracy. It enables the exploration of a subspace without a clear boundary, which we refer to as the soft-constraint subspace. Because of the global search capability, poor initialization still has the opportunity to find the ground state. However, it inevitably gets involved with too many basis states far away from the ground state, leading to excessive resource consumption. Conversely, when the ground state location can be interpolated accurately, a tighter subspace is preferable. In other words, the sub-ansatz discussed in Fig. \ref{fig:figure prodecomp}(b) is more friendly, as it confines the basis states to a smaller region.

We refer to the subspaces spanned by the sub-ansatze as the hard-constraint subspaces. Since each fragment in Eq. \eqref{eq:Dicke linear combination} is also a CCC ansatz, Dicke state $\vert D^n_{n/2}\rangle$ can be recursively decomposed into a linear combination of multiple product Dicke states with amplitude $a_i$ as
\begin{equation}
\begin{split}
\vert D^n_{n/2}\rangle = \sum^{l_u-1}_{i=0} (a_i \bigotimes ^{2^p-1}_{j=0} \vert D_{i,j}\rangle), 
\label{eq:Dicke linear combination recursive}
\end{split}
\end{equation}
where $p$ represents the number of times the recursive decomposition is executed, and $l_u$ is the number of decomposed product states loosely upper bounded by $O(\prod^p_{l=1} (\frac{n}{2^l})^{2^{l-1}})$. Each product state contains $2^p$ product fragments $\vert D_{i,j}\rangle$. This presents a unified recursive equilibrium partition method for general CCC ansatze. For the detailed analysis, refer to Appendix \ref{appen:complexity of RAEP}. In this way, the ansatz for preparing $\vert D^n_{n/2}\rangle$ can be completely partitioned into $\binom{n}{n/2}$ sub-ansatze. Each sub-ansatz corresponds to a certain combination of the $\frac{n}{2}$ $X$ gates. In other words, it leads to an exponential partition complexity in $n$ when the ansatz is partitioned $\lceil$log($n$)$\rceil$ times. Theoretically, more partitions can improve the trainability of VQE on NISQ computers. However, for a quantum algorithm targeting quantum speedups, the number of partitions should be upper bounded.

Intriguingly, convex interpolation can efficiently alleviate the contradiction between speedups and trainability. With convex interpolation, the ground state location can be evaluated recursively in the hard-constraint subspace scenario. Refer to Appendix \ref{appen: Searching in hard-constraint} for a detailed instance. A reasonable analysis shows that the number of sub-ansatze required to be executed is $O(({\rm log}(n))^{{\rm loglog}(n)})$, under loglog($n$) partitions, which has great potential to provide quantum speedups, refer to Appendix \ref{appen:complexity of RAEP}. The crucial aspect lies in accurately evaluating the location of the ground state in each recursion. Otherwise, one misestimation could result in an approximate solution. To alleviate this issue, more efficient interpolation methods are required, which we leave for future work.

The other way round, the approximate solution obtained when the recursive convex interpolation is not precise should be located not far away from the ground state. There should exist a small subspace containing both the approximate solution and the ground state simultaneously. In other words, this subspace effectively bridges the hard-constraint subspaces between the approximate solution and the ground state in the vertical direction. Intriguingly, the local greedy algorithm with the initial basis state closing to $11...100...0$ (the rightmost basis state) can be applied to efficiently search this subspace based on the vertical transfer ability among the partitioned subspaces. Moreover, it is also well-suited for the soft-constraint subspace scenario. In the hybridization manner, the (restricted) global VQE and local greedy algorithms can better leverage their respective advantages. Specifically, VQE searches for the optimal solution on a more global scale. When the obtained solution is an approximate solution, the greedy algorithm provides a significant opportunity to likely transfer the approximate solution to the optimal solution. The hybridization reduces the resource consumption of global search algorithms and enhances the accuracy of local search algorithms. On the contrary, transferring to a better solution prepared by another sub-ansatz implies a potential misestimation of the ground state location. Therefore, the greedy algorithm can be employed as a strategy for estimating the evaluation accuracy of the ground state location.

\section{\label{sec:Experiments}Experiments}

\begin{figure*}[ht]
    \centering
    \includegraphics[width=0.8\textwidth]{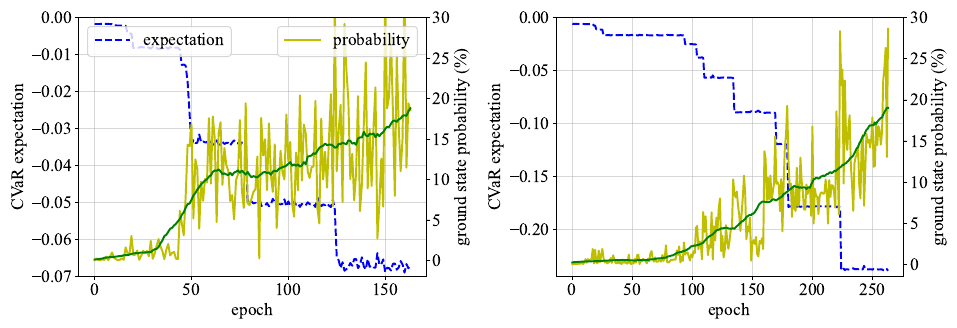}
    \caption{The convergence curves for (a) $\vert D^{12}_6\rangle$ using the soft-constraint subspace method on real dataset, and (b) $\vert D^{40}_{20}\rangle$ using the hard-constraint subspace method on random dataset. The green curve shows the overall ascending trend of the ground state probability. The risk level $q$ is set to $0.9$. The initial confidence level $\alpha$ is set to $0.01$ and is increased to $1.0$ gradually. For the detailed experimental configurations, refer to Table \ref{tab:Experimental configurations} in Appendix \ref{appen: experimental configurations and performance, topology}. In the second experiment, the expectations between epochs 25 to 90 are lower bounded by the soft bottom induced by the correlated parameters. As cloud-based access to the hardware involves queuing and communication, the total running time for the two experiments is $20.43$ minutes (finished on $2024/02/04$ at $22:18:30$) and $70.41$ minutes (finished on $2024/02/20$ at $10:44:34$), respectively. Appendix \ref{appen: Searching in hard-constraint} provides an illustrative example of the procedure used to search for the ground state in the hard-constraint subspace scenario (the second experiment).}
    \label{fig:figure experiment1&2}
\end{figure*}

In this section, we demonstrate the efficiency of the proposed convex interpolation idea for improving the trainability of VQE for solving portfolio optimization and random weighted graph bisection problems. The experiments are performed on the superconducting quantum computer “Wukong” developed by Origin Quantum Computing company, using PyQPanda \cite{Dou2022QPanda}. The configurations of the experiments and the hardware used are detailed in Appendix \ref{appen: experimental configurations and performance, topology}. The conditional value-at-risk (CVaR) \cite{BarkoutsosCVaR2020CVaR} and the iterative-COBYLA \cite{wang2024vqe-po} are utilized as the cost function and the optimization algorithm, respectively.

To mitigate the impact of hardware noise with fewer post processing complexities, only the simplest problem-specific measurement error mitigation \cite{wang2024vqe-po} is utilized. However, this alone is not sufficient to provide adequate performance to guarantee the successful execution of our large-scale experiments. Here we choose to employ the parameter correlation strategy \cite{volkoff2021correlatedparameters,Holmescorrelationinitial}, which is proposed to alleviate the BP problem, to mitigate the impact of hardware noise, and to expedite convergence. Intuitively, the larger the number of correlated parameters, the more errors can be mitigated. However, in the present shallow ansatz scenario, more correlated parameters also lead to a much bumpier cost function landscape (excessively low expressibility), which induces a greater number of spurious local minima. We numerically find that the optimization process would converge to a spurious ground state, which is a superposition state containing the true ground state with a probability that is inversely related to the number of correlated parameters. In other words, the convergence of the CVaR expectation is lower bounded, refer to Appendix \ref{appen:bounded CVaR expectation} for details. Actually, at the beginning of the optimization, more attention should be paid to the rapid descent of the cost expectation. As the parameters are initialized close to the solution, the expectation would descend around the ground state. Convergence to the ground state can be considered in the latter steps of the optimization process. Thus, we decrease the number of correlated parameters gradually during the optimization process to achieve rapid expectation descent and well ground state convergence. As the number of correlated parameters decreases, the search subspace becomes more concentrated and fine-grained.

\subsection{Portfolio optimization experiments}
Two experiments are performed. The first experiment searches for the ground state of $\vert D^{12}_6\rangle$ on real dataset using the soft-constraint subspace method, while the second experiment utilizes $10$ qubits to solve the case of $\vert D^{40}_{20}\rangle$ on random dataset based on the hard-constraint subspace method. The real dataset is downloaded from Tushare (https://tushare.pro/), and the random dataset is generated by $RandomFinanceData$ from Qiskit finance \cite{Qiskit}. The solutions obtained from both experiments are verified and refined by the greedy algorithm.

The convergence curves for both experiments are depicted in Fig. \ref{fig:figure experiment1&2}. The experimental configurations, and the performance and topology of the $12$ superconducting qubits used, are depicted in Appendix \ref{appen: experimental configurations and performance, topology}. In practice, for combinatorial optimization problems, it is unnecessary and unlikely to obtain the ground state with $100\%$ probability. Hence, rapid expectation descent is more critical than well ground state convergence. The confidence level $\alpha$ may be gradually increased from $0.01$ to a value much smaller than $1$, such as $0.1$ or $0.2$, while the correlated parameters can be decreased to an integer larger than $1$. This contributes to a stable convergence process and, additionally, reduces the need for the expressibility of the ansatz. 

Although the bumpy local minima are rearranged to form a much more regular cost function landscape, the stability, rather than the accuracy, of the hardware in the optimization process severely impacts the convergence speed and accuracy. The fluctuation of the curves arises from the nature of iterative-COBYLA, as well as the noise and unstability of the hardware. The ansatz of case $\vert D^{12}_6\rangle$ is composed of $21$ $3$C blocks, while each fragment of case $\vert D^{40}_{20}\rangle$ contains at most $15$ $3$C blocks. The worse crosstalk induced by the parallel execution of the $21$ $3$C blocks leads to the more severe fluctuation of the convergence curve in case $\vert D^{12}_6\rangle$. The experiment are needed to be executed as quickly as possible to promise a stable optimization process.

\subsection{Random weighted graph bisection experiment}

\begin{figure}[ht]
    \centering
    \includegraphics[width=0.42\textwidth]{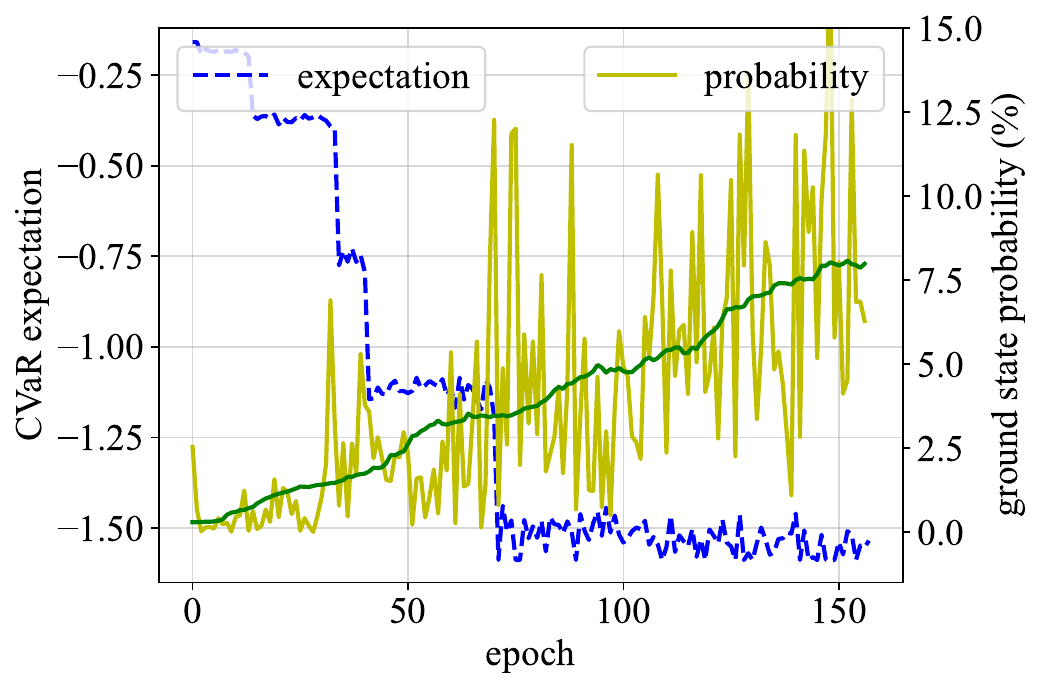}
    \caption{The convergence curve for the random weighted graph bisection experiment $\vert D^{12}_6\rangle$ using the soft-constraint subspace method. The green curve still represents the overall ascending trend of the ground state probability. At this time, the confidence level $\alpha$ is upper bounded by $0.1$. For the detailed experimental configurations, refer to Table \ref{tab:Experimental configurations} in Appendix \ref{appen: experimental configurations and performance, topology}. The total running time for the experiment is $32.48$ minutes (finished on $2024/03/08$ at $14:09:50$).}
    \label{fig:figure experiment3}
\end{figure}

Here, we simply illustrate the scalability of the proposed convex interpolation in solving graph bisection problems using a 12-node ``erdos renyi graph'', which is generated by networkx \cite{networkx}. The probability of adding a new edge for each node is 0.4. Then the weight of each edge is generated using the random seed 123. The nodes are reordered based on the diagonal elements of the Laplacian matrix of the graph. To guarantee the descent of the CVaR expectation, a constant term ``$-20$'' is incorporated into the cost function. To eliminate the symmetry-induced degeneracy of the optimal solution, the highest qubit is fixed to be ``1'' or ``0'' to promise the ground state location is located to the right. In other words, 11 qubits are used. As illustrated in Fig. \ref{fig:figure experiment3}, the expectation descent is retained when the confidence level $\alpha$ reaches the soft cap $0.1$ at about $70$ epochs. Compared to the portfolio optimization case in Fig. \ref{fig:figure experiment1&2}(a), the weight between arbitrary two nodes in the graph is generated randomly. Finally, it is essential to emphasize that reordering based on the diagonal elements of the Laplacian matrix of the graph can not always guarantee the regularity of the cost function landscape. This holds true for the reduced portfolio optimization cost function $C(\bm{\theta}, A)$, as shown in Fig. \ref{fig:figure polyline&distri}(a). The flatter valley is not conducive to expectation descendent. This maybe one of the reasons for the slower ascend of the optimal solution probability. More advanced reordering methods are left for future work.

\section{\label{sec:Conclusion and discussion}Conclusion and discussion}
In this paper, we propose a convex interpolation idea to improve the trainability of VQE on NISQ computers. The ground state location can be predicted, which inspires the proposal of the highly efficient close-to-solution initialization, regular cost function landscape, recursive ansatz equilibrium partition strategies, and a greedy algorithm. This leads to the successful demonstration of solving a $\vert D^{40}_{20}\rangle$ portfolio optimization problem, using $10$ superconducting qubits. To the best of our knowledge, this is the first and the largest portfolio optimization hardware experiment using the distributed quantum computing idea up to now. The convex interpolation method is a universal solution for solving combinatorial optimization problems. Our proposals pave the way for seeking for the quantum advantages in the problem sizes that classical algorithms cannot solve efficiently.

As one of the critical foundations, the interpolation accuracy of the convex curve determines the effectiveness of our algorithm. Interpolating the convex curve using the minimum energies of the $4$ outer sub-ansatze becomes less accurate as the problem sizes increase, especially in the hard-constraint subspace scenario. The fundamental method is to append more interpolation points. However, this necessitates searching for the minimum energies of other sub-ansatze containing many more basis states, making the brute force method unfeasible. For large problem sizes, a machine learning method may have the ability to better predict the distribution characteristics of the convex curve. 
  
The expectation is that the principal component is distributed as smoothly and intensively as possible. In our experiments, the close-to-solution initialization strategy demonstrates good results using merely identical parameter. However, for large enough problem sizes, the smoothness and uniformity of the principal component inevitably deteriorates, particularly in the soft-constraint subspace scenario. In other words, there are several basis states in the principal component owning much greater probabilities than others, rendering the convergence of the algorithm to the local minimum. This is a kind of initialization-induced BP. Hence, more efforts are required to optimize the distribution of the principal component. For example, we can enhance the expressibility of the principal component to explore the cost function landscape on the principal component in a higher-dimensional space. Corresponding to the interpolated ground state location, the $1$ information in the basis states that constitute the principal component are distributed relatively intensively around certain bit areas. Therefore, allocating additional $V_i$s to these areas mainly improves the expressibility of the principal component. This type of local expressibility enhancement (or local overparametrization) strategy can be used to alleviate the impact of BP, hardware noise, and concurrently weaken the bumpy cost function landscape.

The convex interpolation idea serves as a quantum heuristic. It largely eliminates the randomness associated with the ground state location in the proposed VQE-based portfolio optimization. In essence, the inherent data structure of the combinatorial optimization problems closely aligns with this quantum heuristic. This implies that the efficiency of the other classical algorithms can also be improved.

\begin{acknowledgments}
This work was supported by the National Key Research and Development Program of China (Grant No.: 2024YFB4504101).
\end{acknowledgments}

\appendix

\section{\label{appen:Principle analysis of the observation}The observation and its theoretical and numerical verification}

\subsection{Theoretical insight of the observation}

\begin{figure*}[ht]
    \centering
    \includegraphics[width=0.67\textwidth]{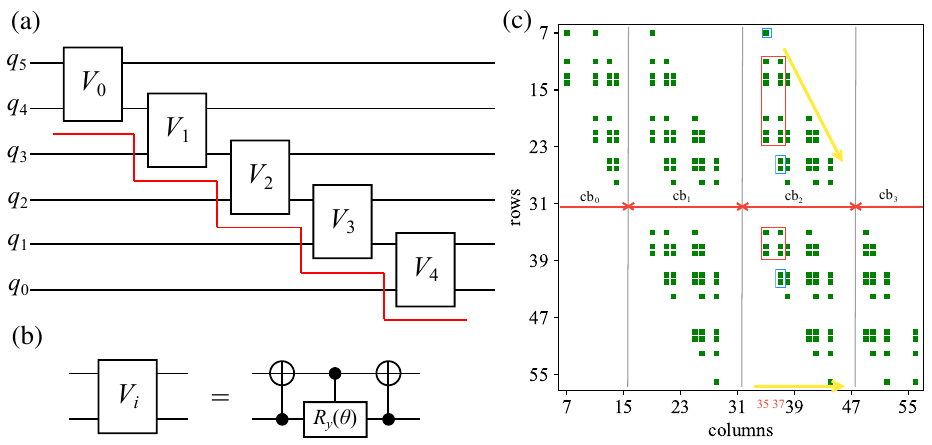}
    \caption{(a) The straight staircase-structure unitary $U_6$ composed of $5$ $2$-qubit unitaries $V_i$. Where $q_5$ is the highest qubit and $q_0$ is the lowest qubit. To produce a Dicke state, an initial state should be prepared before applying $U_n$. (b) The ``CNOT, C-RY, CNOT'' structure of $V_i$, i.e. the $3$C block. The subscript $i$ in $V_i$ represents the parameter $\theta_i$ of the RY gate in $V_i$. (c) The distribution of the $\binom{6}{3}$ basis states in the unitary $U_6$ in the matrix representation; each green square (amplitude) indicates that the corresponding row index represents a $6$-qubit basis state with a Hamming weight of $3$ in its binary representation.}
    \label{fig:figure_A1}
\end{figure*}

Fig. \ref{fig:figure_A1}(a) illustrates the straight staircase-structure unitary $U_n$ by $U_6$ proposed in ref. \cite{wang2024vqe-po}, where $n$ is the number of qubits of unitary $U_n$. The other kinds of structured $U_n$ have similar property. A single layer is used to provide restricted expressibility. This aligns with state-of-the-art analysis results aimed at enhancing the trainability of VQAs. The unitary $U_n$ is synthesized as
\begin{equation}
\begin{split}
U_n = \prod^{0}_{i=n-2}V_i,
\label{eq:Un synthesization}
\end{split}
\end{equation}
\noindent where $V_i$ is the ``CNOT, C-RY, CNOT'' structure shown in Fig. \ref{fig:figure_A1}(b). That is, the matrix representation of $V_i$ used in this paper is
\begin{equation}
V_i=
\begin{pmatrix}
1 &0                 &0                &0 \\
0 &\cos(\theta_i/2)   &\sin(\theta_i/2)  &0 \\
0 &-\sin(\theta_i/2)  &\cos(\theta_i/2)  &0 \\
0 &0                 &0                &1
\end{pmatrix},
\label{eq:matrix representation of Vi}
\end{equation}
where $\theta_i$ is the parameter belonging to $[0,\pi]$. For the simplest case of $n=2$, $U_2=V_0$, corresponding to the $V_0$ part on qubits $q_5,q_4$ in Fig. \ref{fig:figure_A1}(a). The initial state can be prepared as $\vert 01\rangle$ or $\vert 10\rangle$, corresponding to columns $1$ or $2$ of $U_2$. Then the Dicke state $\vert D^2_1\rangle$ can be prepared by $\vert D^2_1\rangle = U_2\vert 01\rangle$ or $\vert D^2_1\rangle = U_2\vert 10\rangle$. Here, we focus on the completeness of the basis state with corresponding Hamming weight and ignore the differences of the probability amplitudes of $\vert D^2_1\rangle$. In consequence, the above two Dicke states $\vert D^2_1\rangle$ are casually regarded as the same Dicke state. In addition, Dicke states $\vert D^2_0\rangle$ and $\vert D^2_2\rangle$ can be prepared by initializing to $\vert 00\rangle$ and $\vert 11\rangle$ respectively, which will be ignored in the following analysis.

For case of $n=4$, $U_4$ is synthesized by $(I\otimes V_2)(I\otimes V_1\otimes I)(V_0\otimes I)$ as
\setcounter{MaxMatrixCols}{16}  % 设置矩阵可以显示的最大列数
\begin{widetext}
\begin{equation}
U_4=
\scalebox{0.8}{$
\begin{pmatrix}
1&0&0&0&0&0&0&0&0&0&0&0&0&0&0&0\\
0&c_2&s_2c_1&0&s_2s_1c_0&0&0&0&s_2s_1s_0&0&0&0&0&0&0&0\\
0&-s_2&c_2c_1&0&c_2s_1c_0&0&0&0&c_2s_1s_0&0&0&0&0&0&0&0\\
0&0&0&c_1&0&s_1c_0&0&0&0&s_1s_0&0&0&0&0&0&0\\
0&0&-s_1&0&c_1c_0&0&0&0&c_1s_0&0&0&0&0&0&0&0\\
0&0&0&-c_2s_1&0&c_2c_1c_0&s_2c_0&0&0&c_2c_1s_0&s_2s_0&0&0&0&0&0\\
0&0&0&s_2s_1&0&-s_2c_1c_0&c_2c_0&0&0&-s_2c_1s_0&c_2s_0&0&0&0&0&0\\
0&0&0&0&0&0&0&c_0&0&0&0&s_0&0&0&0&0\\
0&0&0&0&-s_0&0&0&0&c_0&0&0&0&0&0&0&0\\
0&0&0&0&0&-c_2s_0&-s_2c_1s_0&0&0&c_2c_0&s_2c_1c_0&0&s_2s_1&0&0&0\\
0&0&0&0&0&s_2s_0&-c_2c_1s_0&0&0&-s_2c_0&c_2c_1c_0&0&c_2s_1&0&0&0\\
0&0&0&0&0&0&0&-c_1s_0&0&0&0&c_1c_0&0&s_1&0&0\\
0&0&0&0&0&0&s_1s_0&0&0&0&-s_1c_0&0&c_1&0&0&0\\
0&0&0&0&0&0&0&c_2s_1s_0&0&0&0&-c_2s_1c_0&0&c_2c_1&s_2&0\\
0&0&0&0&0&0&0&-s_2s_1s_0&0&0&0&s_2s_1c_0&0&-s_2c_1&c_2&0\\
0&0&0&0&0&0&0&0&0&0&0&0&0&0&0&1 \\
\end{pmatrix}$},
\label{eq:matrix representation of U4}
\end{equation}
\end{widetext}
where $\cos(\frac{\theta_i}{2})$ and $\sin(\frac{\theta_i}{2})$ are abbreviated to $c_i$ and $s_i$ respectively. The corresponding circuit can be illustrated as the part composed by $V_0$, $V_1$, and $V_2$ on qubits $q_5,q_4,q_3,q_2$ in Fig. \ref{fig:figure_A1}(a). As can be seen in Eq. \eqref{eq:matrix representation of U4}, the Hamming weight of the elements of column $j$ of unitary $U_4$ equals to the Hamming weight of $j$. For instance, by setting the initial state to be $\vert 0101\rangle$, the elements of column $5$ of $U_4$ is prepared to be
\begin{equation}
\begin{split}
\vert \psi_5\rangle=&U_4\vert 0101\rangle=s_1c_0\vert 0011\rangle + c_2c_1c_0\vert 0101\rangle \\
&- s_2c_1c_0\vert 0110\rangle - c_2s_0\vert 1001\rangle + s_2s_0\vert 1010\rangle,
\label{eq:psi 5}
\end{split}
\end{equation}
where the Hamming weights of the input state and the basis states of the output state are all $2$. However, the basis states with Hamming weight $2$ on $\vert \psi_5\rangle$ is not complete, missing basis state $\vert 1100\rangle$. And there are overlap and non-overlap elements among different columns with the same Hamming weight. The completeness of the basis states with the target Hamming weight can be guaranteed by preparing the initial state as a linear combination of certain columns with the same Hamming weight. In fact, the linear combination of the columns could improve the expressibility of the ansatz.

To display matrix $U_n$ more intuitively for cases with larger $n$, the elements are represented by the green squares, as illustrated in Fig. \ref{fig:figure_A1}(c). Intriguingly, the distribution is very regular. The matrix can be divided into $4$ column blocks, cb$_0$ to cb$_3$, separated by the vertical gridlines, corresponding to columns $0-15$, $16-31$, $32-47$, and $48-63$, respectively. In cb$_2$, the following observations can be found: (i) In each column, the ``$1$''s information of the basis states are distributed relatively intensively around certain bit areas. The areas gradually transfer from low bit to high bit as the column index increases, as the yellow arrows indicated. (ii) As mentioned above, for any two columns, their contained elements can be divided into two regions: the overlap region and the non-overlap region. As illustrated in Fig. \ref{fig:figure_A1}(c), the overlap and non-overlap regions of columns $35$ and $37$ of the staircase-structure $U_6$ are depicted in the red and blue boxes, respectively. It shows that the basis states of the two columns in the non-overlap region own large Hamming distances. Concurrently, other states that have much smaller Hamming distances between each other are categorized into the overlap region. Additionally, as the Hamming distance between two columns gradually increases, their overlap region fades away.

\begin{figure*}[ht]
    \centering
    \includegraphics[width=0.67\textwidth]{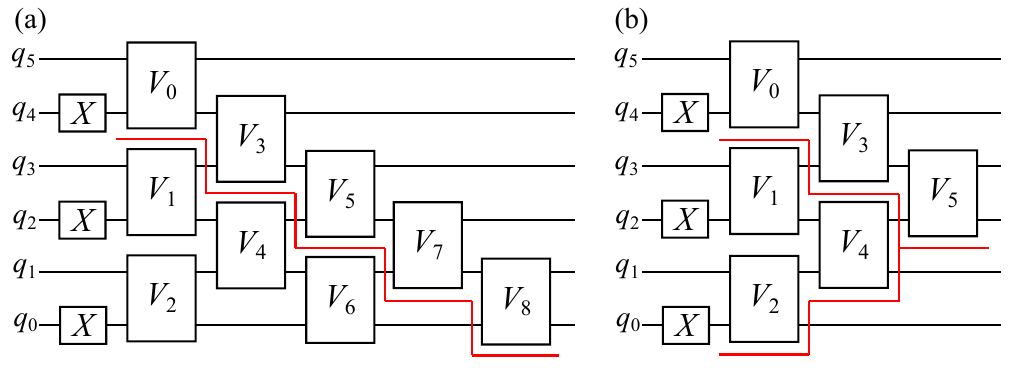}
    \caption{The CCC ansatze for preparing Dicke state $\vert D^6_3\rangle$ using (a) the straight staircase $U_6$, and (b) the folded staircase $U_6$.}
    \label{fig:figure_A2}
\end{figure*}

Inspired by the observation that the energy of a state with a larger Hamming distance from the ground state results in a greater energy gap away from the ground state energy, the overlapped clustering property should be able to guarantee a flatter valley, interpolated by the minimum energies of the states within the columns. The ground state would be located in more than one sub-ansatz when it distributes in an overlap region. Otherwise, the flatter valley can still promise a smaller difference between the ground state energy and the minimum energies of the sub-ansatze adjacent to the sub-ansatz containing the ground state. In other words, an approximate solution very close to the ground state can be guaranteed. However, the existence of the overlap region also increases the search space dimension. Apparently, the clustering property can be extended to encompass multiple columns that form a column set based on the Hamming distance between these columns.

\begin{figure*}[ht]
    \centering
    \includegraphics[width=0.6\textwidth]{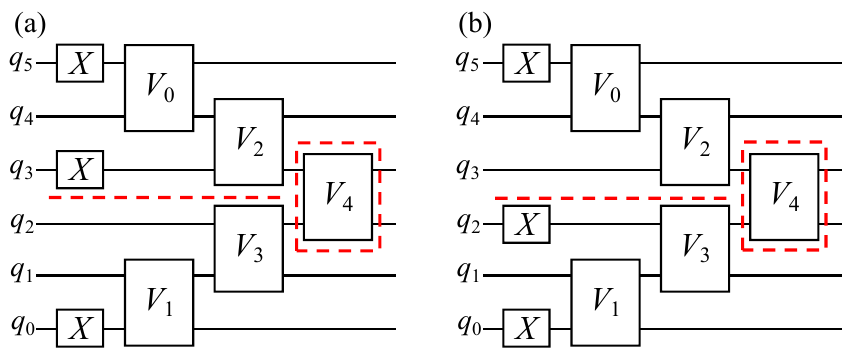}
    \caption{The decomposition of the folded CCC ansatz in ref. \cite{wang2024vqe-po}, as shown in Fig. \ref{fig:figure_A2}(b). Firstly, the block for preparing $\vert D^2_1\rangle$ on $q_3$ and $q_2$ are decomposed as $\vert D^2_1\rangle = \sum^1_{i=0}a_i\vert D^1_i\rangle\otimes\vert D^1_{1-i}\rangle$, as detailed in Appendix \ref{appen:Proof product decomposition of CCC ansatz}. This yields the two sub-ansatze. Secondly, in each sub-ansatz, the $2$-qubit unitary $V_4$ in the red dashed box is cut off and simulated classically. The remaining two product fragments, separated by the red dashed line, can be executed by $3$ qubits. As can be seen, due to the entanglement of $V_4$ between the two fragments, there is an overlap region between the spaces spanned by the two sub-ansatze. This results in a flatter convex curve.}
    \label{fig:figure_A3}
\end{figure*}

The remaining task is to find an efficient way to construct several suitable column sets to achieve the expectedly regular distribution of the minimum energies. In fact, there are plenty of structures that can prepare the same Dicke state with different amplitude configurations. As illustrated in Fig. \ref{fig:figure_A2}, both ansatze can prepare the Dicke state $\vert D^6_3\rangle$ that contain all $\binom{6}{3}$ basis states. However, they have different staircase structures and hence different amplitude configurations. The equal partition of the ansatz, as shown in Fig. \ref{fig:figure_A2}(b) summarized in ref. \cite{wang2024vqe-po} (see Fig. \ref{fig:figure_A3}), motivates the rigorous derivation of the theoretical results in Appendix \ref{appen:Proof product decomposition of CCC ansatz}. Clustering the basis states in this way achieves several advantages, as depicted in Fig. \ref{fig:figure prodecomp} in the main text. (i) There are no overlap regions, eliminating the redundant sampling complexity. (ii) Each sub-ansatz is constructed by several fragments with the same number of qubits in a tensor product manner. This type of circuit can be efficiently simulated by a smaller quantum computer. From a different perspective, the CCC ansatze shown in Fig. \ref{fig:figure_A2} can be regarded as extracting the elements stored in column $010101$ of the $6$-qubit unitary constructed by all of the $2$-qubit unitary $V_i$s. In other words, column $010101$ contains all $\binom{6}{3}$ basis states.

\subsection{Theoretical and numerical verification}

% b与x的对应关系需要调整一下，x是由多个二进制变量组成的矢量，而b是个二进制串矢量，也就是他们的变量数不同，x的变量数为n，而b的变量数为1；
This section presents the theoretical analysis and extensive numerical simulations to verify the correctness and effectiveness of our intuition, namely the convex interpolation idea. In Eq. \eqref{eq:mean-variance}, $\bm{x}$ is a set of vectors, where each vector consists of $n$ binary variables $x_i$. Meanwhile $\bm{b}=\{b_0,b_1,...,b_{2^n-1}\}$, with the basis state $b_i=b_{i,0}b_{i,1}…b_{i,n-1}$, represents the corresponding set of binary string variable. Here the binary variable is replaced with the binary string variable. According to Eq. \eqref{eq:mean-variance}, the convexity of the curve is determined by two factors: the expected return of any single asset, $\bm{\mu}$, and the covariance between any two assets, $A$. As the expected returns are independent to each other, we can strictly prove the convexity of $-\bm{\mu^T b}$ (equivalent to $-\bm{\mu^T x}$) theoretically. However, $\bm{b^T}A \bm{b}$ cannot be guaranteed to be convex, which is numerically simulated to demonstrate its trivial violation to our intuition.

At first, we assume that $\mu_i\neq \mu_j$ for $i,j \in \{0,1,2,…,n-1\}$ with $i\neq j$, and that the ground state is located in subspace sa$_k$, as referenced in Eq. \eqref{eq:Dicke linear combination}. Then the minimum energy of subspace sa$_{k-1}$ can be obtained by exchanging the asset with the largest return in the left $\frac{n}{2}$-qubit fragment, $\mu_{l1}$, with the asset who owns the smallest return in the right, $\mu_{r1}$. The gradient between subspaces sa$_k$ and sa$_{k-1}$ can be evaluated as
\begin{equation}
\begin{split}
g_1 = \frac{E_k-E_{k-1}}{k-(k-1)}= E_g-E_{k-1} = -\mu_{r1}+\mu_{l1}.
\label{eq:gradient 1}
\end{split}
\end{equation}
Similarly, the minimum of subspace sa$_{k-2}$ can be obtained by exchanging the two assets who own the largest and the second largest returns in the left fragment with the two assets who own the smallest and the second smallest returns in the right. It is evident that the four exchanged assets are $\mu_{l1}, \mu_{l2}$ and $\mu_{r1},\mu_{r2}$, where $\mu_{l1}$ and $\mu_{r1}$ are the same as in Eq. \eqref{eq:gradient 1}. The gradient between subspaces sa$_{k-1}$ and sa$_{k-2}$ is
\begin{equation}
\begin{split}
g_2 = & \frac{E_{k-1}-E_{k-2}}{(k-1)-(k-2)}= E_{k-1} - E_{k-2} \\
= & (E_g-(-\mu_{r1} +\mu_{l1}))- (E_g \\
&-(-\mu_{r1} - \mu_{r2} + \mu_{l1} + \mu_{l2})) \\
= & -\mu_{r2}+\mu_{l2}.
\label{eq:gradient 2}
\end{split}
\end{equation}
As assumed above, the order of the asset from small to large is $…$, $\mu_{l3}$, $\mu_{l2}$, $\mu_{l1}$, $\mu_{r1}$, $\mu_{r2}$, $\mu_{r3}$, $…$. We obtain $g_2 = -\mu_{r2}+\mu_{l2} < -\mu_{r1}+\mu_{l1} < 0$. Similarly, it can be proven that the term $-\bm{\mu^T b}$ is strictly convex with respect to the minimum energies of each subspace. That is to say, when $q = 0$, the cost function strictly follows our intuition.

\begin{figure*}[ht]
    \centering
    \includegraphics[width=0.8\textwidth]{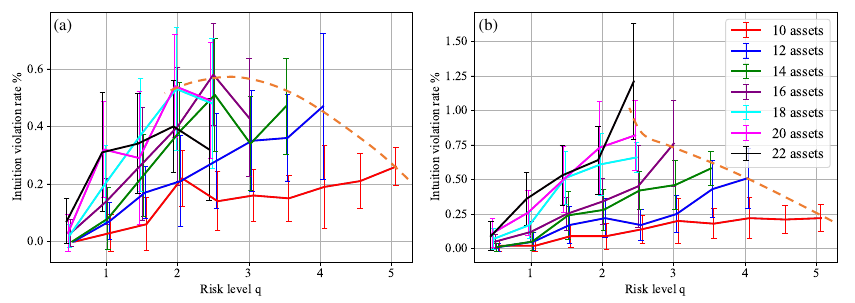}
    \caption{The average violation rate and its variance for (a) the real dataset and (b) the random dataset. The asset pools contain $10-22$ assets, with the risk value $q$ ranging from $0.5$ to $5$ in steps of $0.5$. The curves are adjusted by $0.01(16-i)$, where $i$ is the number of assets, to better highlight the variances. As shown in the theoretical analysis above, the violation rate is $0$ when $q=0$.}
    \label{fig:figure_realdataset}
\end{figure*}

However, the term $\bm{b^TA b}$ may violate our intuition when the exchanged assets exhibit negative correlations. Then for the entire cost function $q\bm{x}^TA\bm{x}-\bm{\mu}^T\bm{x}$, we derive two new intuitions. \textbf{Intuition 1}: As the number of assets increases, the violation rate becomes higher. \textbf{Intuition 2}: As the risk level $q$ increases, the number of experiments yielding a negative ground state energy decreases. The experimental results of the brute-force numerical simulations, conducted on the order of $\bm{10^7}$ asset pools, are depicted in Fig. \ref{fig:figure_realdataset}. The intuition violation refers to changes in the magnitude relationship between the minimum energies of adjacent subspaces. For instance, changing from $E_k > E_{k-1}$ to $E_k \leq E_{k-1}$. Each curve corresponds to a specific number of assets, ranging from $10$ to $22$. Each simulation is averaged over $10$ experiments based on either a real dataset—Tushare (https://tushare.pro/, which contains $785$ stocks from China A-shares)— or random dataset from Qiskit finance \cite{Qiskit}. In each experiment, $1000$ asset pools are selected from $20000$ configurations. As $q$ increases, the number of asset pools with a negative ground state energy decreases rapidly. And when $1000$ cases with a negative ground state energy can no longer be selected, the simulation terminates. As depicted in the figure, as the number of qubits increases, the curve terminates at progressively smaller values of $q$, which follows intuition 1.

Additionally, the violation rate of larger asset pools increases more rapidly with $q$ than that of smaller ones, as more assets implies more negative correlations contributing to the violation (intuition 1). However, the violation rate of the real dataset begins to decrease when $q$ reaches a certain threshold (the peak of each curve). The dashed yellow represents a fit for these thresholds. As the number of asset pools exceeds $18$, the violation rate starts to decrease. This may stem from intuition 2, suggesting that to ensure a meaningfully negative ground state energy, the negative correlations between assets in a real asset pool with a larger $n$ are neutralized in some way. More violation cases are transferred to the orthogonal subspace with a positive ground state energy. For large asset pools and high values of $q$, we observe that the violation rate of the random dataset spikes, differing significantly from that of the real dataset. Additionally, for small asset pools, the violation rate increases approximately linearly. This may result from the amplification of inaccurate correlations between random assets due to a large value of $q$.

% \vspace{5pt}

\section{\label{appen:Proof product decomposition of CCC ansatz}Product decomposition of Dicke states}

Lemma $1$ of ref. \cite{bartschi2019deterministic} states that the equal-superposition Dicke state $\vert D^n_k\rangle$ has the following inductive sum form
\begin{equation}
\begin{split}
\vert D^n_k\rangle = \sqrt{\frac{k}{n}}\vert 1\rangle\vert D^{n-1}_{k-1}\rangle + \sqrt{\frac{n-k}{n}}\vert 0\rangle\vert D^{n-1}_{k}\rangle. \label{eq:decomposition of Dicke state}
\end{split}
\end{equation}
In this paper, the major focus is on the completeness of the basis states of a Dicke state. Hence, we ignore the amplitudes at first and proof that the highly entangled Dicke state $\vert D^n_{n/2}\rangle$ can be decomposed into ($\frac{n}{2}+1$) product states that are composited of two $\frac{n}{2}$-qubit complementary Dicke states. That is, the sum of the ``$1$''s in the two $\frac{n}{2}$-qubit Dicke states is $\frac{n}{2}$. In the following, we use the notation ``$=_c$'' to represent the basis state \textbf{$\bm{c}$}ompleteness equivalence (The subscript ``$c$'' represents ``completeness''). Our proof uses induction method.

Firstly, we prove the generalized cases of decomposing an arbitrary Dicke state $\vert D^n_k\rangle$ into a linear combination of $2$ product fragments. In this paper, we assume $k$ is an integer not greater than $n/2$. Cases with $k > \frac{n}{2}$ can be converted using the equivalent substitution relation of $\vert D^n_k\rangle=X^{\otimes n}\vert D^n_{n-k}\rangle$. Based on Eq. \eqref{eq:decomposition of Dicke state}, the following induction can be obtained:

(1) For cases satisfying $j \leq k$ where $j$ is the number of qubits partitioned into the upper fragment. Without lose of generality, $j$ is also assumed to be an integer not greater than $\frac{n}{2}$:
\begin{equation}
\begin{split}
\vert D^n_k\rangle =_c& \vert D^1_1\rangle\vert D^{n-1}_{k-1}\rangle + \vert D^1_0\rangle\vert D^{n-1}_k\rangle \\
=_c& \sum^1_{i=0}\vert D^1_i\rangle \otimes\vert D^{n-1}_{k-i}\rangle, \label{eq:induction no larger 0}
\end{split}
\end{equation} %%%%%%%%%%%%%
\begin{equation}
\begin{split}
\vert D^n_k\rangle =_c \ & \vert D^1_1\rangle(\vert D^1_1\rangle\vert D^{n-2}_{k-2}\rangle + \vert D^1_0\rangle\vert D^{n-2}_{k-1}\rangle) \\
&+ \vert D^1_0\rangle(\vert D^1_1\rangle\vert D^{n-2}_{k-1}\rangle + \vert D^1_0\rangle\vert D^{n-2}_k\rangle) \\
=_c \ & \vert D^2_2\rangle\vert D^{n-2}_{k-2}\rangle + \vert D^2_1\rangle\vert D^{n-2}_{k-1}\rangle + \vert D^2_0\rangle\vert D^{n-2}_k\rangle \\
=_c \ & \sum^2_{i=0}\vert D^2_i\rangle\otimes\vert D^{n-2}_{k-i}\rangle,
\label{eq:induction no larger 1}
\end{split}
\end{equation} %%%%%%%%%%%%%
\hspace{4cm}$......$
\begin{equation}
\begin{split}
\vert D^n_k\rangle =_c \sum^{j-1}_{i=0}\vert D^{j-1}_i\rangle\otimes\vert D^{n-(j-1)}_{k-i}\rangle, \label{eq:induction no larger k-1}
\end{split}
\end{equation} %%%%%%%%%%%%%
\begin{equation}
\begin{split}
\vert D^n_k\rangle =_c  & \sum^{j-1}_{i=0}\vert D^{j-1}_i\rangle\otimes\vert D^{n-(j-1)}_{k-i}\rangle \\
=_c & \sum^{j-1}_{i=0} \vert D^{j-1}_i\rangle \otimes(\vert D^1_1\rangle\vert D^{n-j}_{k-i-1}\rangle + \vert D^1_0\rangle\vert D^{n-j}_{k-i}\rangle) \\
=_c  & \sum^{j-1}_{i=0}(\vert D^j_{i+1}\rangle\vert D^{n-j}_{k-i-1}\rangle + \vert D^j_i\rangle\vert D^{n-j}_{k-i}\rangle) \\
=_c &(\sum^{j-1}_{i=0}\vert D^j_i\rangle\otimes\vert D^{n-j}_{k-i}\rangle) + \vert D^j_j\rangle\vert D^{n-j}_{k-j}\rangle \\
=_c  & \sum^{j}_{i=0}\vert D^j_i\rangle\otimes\vert D^{n-j}_{k-i}\rangle.
\label{eq:induction no larger k}
\end{split}
\end{equation}

(2) For cases satisfying $j > k$. As can be seen in Eq. \eqref{eq:induction no larger k}, the upper fragment contains ($j+1$) terms, i.e. $\vert D^j_0\rangle$, $\vert D^j_1\rangle$, $...$, $\vert D^j_j\rangle$. However, for $j > k$, these terms reduce to $\vert D^j_0\rangle$, $\vert D^j_1\rangle$, $...$, $\vert D^j_k\rangle$. That is to say, the Dicke state can be decomposed as
\begin{equation}
\begin{split}
\vert D^n_k\rangle =_c \sum^{k}_{i=0}\vert D^k_i\rangle\otimes\vert D^{n-k}_{k-i}\rangle, \label{eq:induction larger 0}
\end{split}
\end{equation} %%%%%%%%%%%%%
\begin{equation}
\begin{split}
\vert D^n_k\rangle =_c  & \sum^{k}_{i=0}\vert D^k_i\rangle\otimes\vert D^{n-k}_{k-i}\rangle \\
=_c &\sum^{k-1}_{i=0} \vert D^k_i\rangle \otimes\vert D^{n-k}_{k-i}\rangle + \vert D^k_k\rangle\otimes\vert D^{n-k}_0\rangle \\
=_c  & (\sum^{k-1}_{i=0}\vert D^k_i\rangle\otimes(\vert D^1_1\rangle\vert D^{n-k-1}_{k-i-1}\rangle + \vert D^1_0\rangle\vert D^{n-k-1}_{k-i}\rangle)) \\
& + \vert D^k_k\rangle\otimes\vert D^1_0\rangle\vert D^{n-k-1}_0\rangle \\
=_c  & (\sum^{k-1}_{i=0}\vert D^{k+1}_i\rangle\otimes\vert D^{n-k-1}_{k-i}\rangle) + \vert D^{k+1}_k\rangle\vert D^{n-k-1}_0\rangle \\
=_c  & \sum^{k}_{i=0}\vert D^{k+1}_i\rangle\otimes\vert D^{n-k-1}_{k-i}\rangle,
\label{eq:induction larger 1}
\end{split}
\end{equation} %%%%%%%%%%%
\hspace{5cm}$......$
\begin{equation}
\begin{split}
\vert D^n_k\rangle =_c \sum^k_{i=0}\vert D^{j-1}_i\rangle\otimes\vert D^{n-j+1}_{k-i}\rangle, \label{eq:induction larger k-1}
\end{split}
\end{equation} %%%%%%%%%%%%%
\begin{equation}
\begin{split}
\vert D^n_k\rangle =_c  & \sum^k_{i=0}\vert D^{j-1}_i\rangle\otimes\vert D^{n-j+1}_{k-i}\rangle \\
=_c  & (\sum^{k-1}_{i=0}\vert D^{j-1}_i\rangle\otimes\vert D^{n-j+1}_{k-i}\rangle) + \vert D^{j-1}_k\rangle\vert D^{n-j+1}_0\rangle \\
=_c  & (\sum^{k-1}_{i=0}\vert D^j_i\rangle\otimes\vert D^{n-j}_{k-i}\rangle) + \vert D^j_k\rangle\vert D^{n-j}_0\rangle \\
=_c  & \sum^{k}_{i=0}\vert D^j_i\rangle\otimes\vert D^{n-j}_{k-i}\rangle.
\label{eq:induction larger k}
\end{split}
\end{equation}
The Dicke state $\vert D^n_k\rangle$ can be regarded as a linear combination of multiple product states with amplitudes $a_i$ as
\begin{equation}
\begin{split}
\vert D^n_k\rangle = \sum^{\operatorname{min}\{j,k\}}_{i=0}a_i\vert D^j_i\rangle\otimes\vert D^{n-j}_{k-i}\rangle, \label{eq:induction sum up}
\end{split}
\end{equation} %%%%%%%%%%%%%
where the product decomposition can only be guaranteed with specific amplitude configurations. For instance, the product decomposition of $\vert D^4_2\rangle = a_0\vert0011\rangle + a_1\vert0101\rangle + a_2\vert0110\rangle + a_3\vert1001\rangle + a_4\vert1010\rangle + a_5\vert1100\rangle$ should satisfy $a_1/a_2 = a_3/a_4$. Apparently, identical amplitude belongs to one of these configurations.

Hence, to decompose the highly entangled Dicke state $\vert D^{n}_{n/2}\rangle$ into a linear combination of $2$ equal-size fragments, Eq. \eqref{eq:induction sum up} is reformulated as
\begin{equation}
\begin{split}
\vert D^n_{n/2}\rangle = \sum^{n/2}_{i=0}a_i\vert D^{n/2}_i\rangle\otimes\vert D^{n/2}_{n/2-i}\rangle. \label{eq:induction sum up reformulated}
\end{split}
\end{equation} %%%%%%%%%%%%%
This is Eq. \eqref{eq:Dicke linear combination} in the main text. To sum up, we can partition the CCC ansatz for preparing $\vert D^n_{n/2}\rangle$ into $\frac{n}{2}+1$ product sub-ansatze. Then the $n$-qubit Dicke state can be simulated using a $\frac{n}{2}$-qubit quantum computer. The total sampling complexity is $O(n)$. While the circuit cutting technique introduces an exponential sampling complexity of $O(4^n)$ \cite{Harada2023optimalcircuitcuttingwithout}, which completely inundates the potential quantum advantages.

Clearly, the fragments $\vert D^{n/2}_i\rangle$ and $\vert D^{n/2}_{n/2-i}\rangle$ in Eq. \eqref{eq:induction sum up reformulated} are also complete Dicke states. Therefore, a Dicke state can be decomposed recursively into multiple equilibrium or non-equilibrium fragments. Recursively, the obtained product terms $\vert D^k_i\rangle$ and $\vert D^k_{k-i}\rangle$ are also Dicke states which can be decomposed further. This leads to the proposal of the unified recursive equilibrium partitioning method. It is important to emphasize that the decomposability of a Dicke state is restricted to specific amplitude configurations.

\section{\label{appen: simulability proves}Simulability of Dicke state and its decompositions}
In practice, a Dicke state prepared by the ansatz (and by fragments of the sub-ansatze) during the optimization process is typically far from a state that can be decomposed into the product form of Eq. \eqref{eq:induction sum up}. As mentioned at the beginning, the main focus is on the completeness of the basis states of a Dicke state, rather than on the normalization of the amplitudes. In other words, the decomposition implicitly enhances the expressibility of the CCC ansatze. Moreover, this also renders the optimization process infeasible to simulate efficiently on a classical computer \cite{vidalefficsimu}.

Here we also assume $k \leq n/2$, since any case with $k > n/2$ can be converted to $n-k \leq n/2$. Based on Stirling's approximation, the binomial coefficient $\binom{n}{n/2}$ has the same asymptotic complexity as $\frac{2^n}{\sqrt{\pi n/2}}$, i.e. $\binom{n}{n/2} = O(\frac{2^n}{\sqrt{n}})$. This implies exponential complexity with respect to 
$n$. Therefore, for an arbitrary Dicke state $\vert D^n_{n/2}\rangle$, the size of the corresponding subspace grows as $O(2^n)$ for $n$ qubits/assets.

Obviously, any Dicke state $\vert D^n_k\rangle$  with sufficiently small (or large) $k$ can be efficiently simulated classically. For instance, the minimum energies of the four outer subspaces can be obtained efficiently using classical methods, due to their small values of $k$ (e.g., $k=1$ and $k=2$). Here the classical simulation efficiency is assumed to be upper bounded by polynomial time (or resources) used. The binomial coefficient $\binom{n}{k}=\frac{n(n-1)...(n-k+1)}{k!}$ , with small $k = O(logn)$, can be asymptotically upper bounded by the super-polynomial complexity $\frac{n^k}{k!}=O(n^{logn})$. Hence, cases with a $k > O(logn)$ cannot be efficiently simulated classically.

For the cases where $|\frac{n}{2}-k|=O(logn)$, we note that analyzing $k \leq n/2$ and $k> n/2$ yields the same complexity,
\begin{equation}
\begin{split}
r=&\binom{n}{n/2}/\binom{n}{n/2-logn}\\
=&\frac{(n/2+logn)...(n/2+1)}{n/2...(n/2-logn+1)}. 
\label{eq:ratio to the most comp}
\end{split}
\end{equation} %%%%%%%%%%%%%
The ratio lies within the following bounds, 
\begin{equation}
\begin{split}
\prod_{logn}&\frac{n+2logn}{n}<r=\frac{n+2logn}{n}\frac{n+2(logn-1)}{n-2}\\
&...\frac{n+2}{n-2(logn-1)}<\prod_{logn}\frac{n+2}{n-2(logn-1)}. 
\label{eq:ratio interval}
\end{split}
\end{equation} %%%%%%%%%%%%%
Based on the Taylor series expansion  of logarithmic function, $lim_{x\rightarrow0}log(1+x)=x$, and omitting the constant factor $1/ln2$, when $n$ is large enough, the logarithm of the inequality can be simplified as
\begin{equation}
\begin{split}
\frac{2log^2n}{n}<logr<\frac{2log^2n}{n+2-2logn}. 
\label{eq:ratio simplified}
\end{split}
\end{equation} %%%%%%%%%%%%%
Therefore, the complexity of $\binom{n}{n/2-logn}$ is asymptotically $O(\frac{2^{n-2log^2n/n}}{\sqrt{n}})$. This can still be regarded as exponential complexity.

Additionally, in the partitioned scenario, the complexity of each fragment determines its simulability. To ensure that the $i$th partition with $n/2^i$ qubits cannot be efficiently simulated classically, the value of $k$ must be greater than $log(n/2^i)$. For the case of $loglogn$ partitions, as summarized in Appendix \ref{appen:complexity of RAEP}, the most complex fragment(s) correspond to the state $\vert D^{n/logn}_{n/(2logn)}\rangle$. Its complexity is asymptotically $\frac{2^{n/logn}}{\sqrt{\pi n/(4logn)}}$, which is subexponential in $n$. In practice, when $n/logn$ exceeds the threshold for efficient classical simulation, say for $n \geq 450$, simulating a fragment of the form $\vert D^{n/logn}_{n/(2logn)}\rangle$ also becomes exponentially difficult. 

Some studies \cite{anschuetz2023efficient} have shown that Dicke states can be efficiently simulated classically when they are symmetric—that is, invariant under qubit permutations. For example, an equal superposition Dicke state is symmetric and can therefore be simulated efficiently on a classical computer. In contrast, the variational parameters in the present scenario typically produce asymmetric variational Dicke states, making efficient classical simulation a low-probability event. In summary, for a Dicke state $\vert D^n_k\rangle$, whether prepared by a complete ansatz or a fragment, classical simulation is asymptotically feasible when $k \leq O(logn)$, while cases with $k> O(logn)$ require quantum execution. In the latter case, efficient classical simulation can be ruled out.

\section{\label{appen: completeness and reachability proves}Completeness and reachability of CCC ansatze}
\begin{figure*}[ht]
    \centering
    \includegraphics[width=0.9\textwidth]{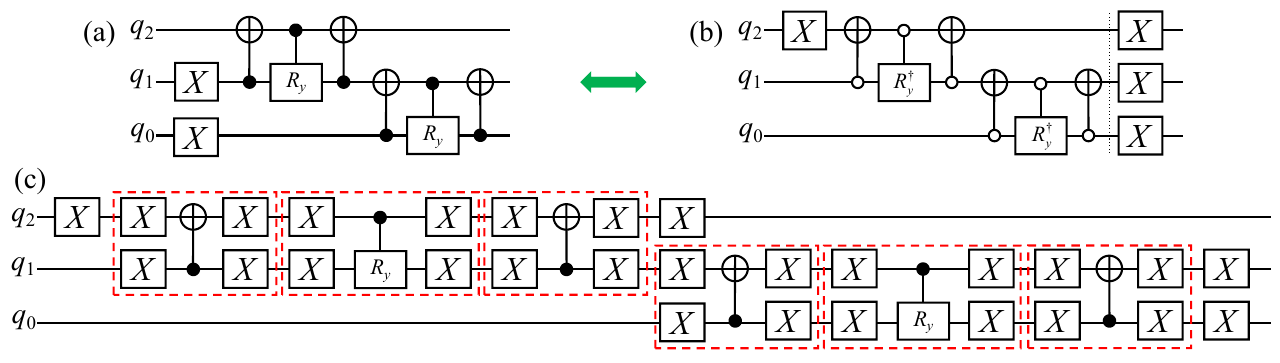}
    \caption{(a) The CCC ansatz for directly preparing $\vert D^3_2\rangle$. The $X$ gates are used to input $1$ information into different 3C blocks. (b) The transformed circuit. As analyzed in Appendix \ref{appen:Principle analysis of the observation}, input $\vert 01\rangle$ or $\vert 10\rangle$ produces the same basis states. Therefore, locating the first $X$ gate on $q_2$ or $q_1$ does not impact the completeness of the produced basis states. (c) The transformation method. The identity gate $I=XX$ is inserted into qubit segments connected to CNOT and C-RY gates. The 3C block $V_i$ in Eq. \eqref{eq:matrix representation of Vi} is transformed to $V^{\dagger}_i$.}
    \label{fig: circuit identity}
\end{figure*}

The completeness means the state prepared by a CCC ansatz is a complete Dicke state, as shown in Eq. \eqref{eq:Dicke state definition}. While, in the present scenario, the reachability represents that any basis state prepared by the CCC ansatze can be obtained with $100\%$ probability when the parameters are set to specific values. In Appendix \ref{appen:Proof product decomposition of CCC ansatz}, we assume $k$ of $\vert D^n_k\rangle$ is not greater than $\frac{n}{2}$. And the cases with $k>\frac{n}{2}$ can be obtained by the equivalent substitution relation of $\vert D^n_k\rangle=X^{\otimes n}\vert D^n_{n-k}\rangle$. Before the proof, we give the rationality of this relation by taking $\vert D^3_2\rangle$ as a warm-up example. As depicted in Fig. \ref{fig: circuit identity}, the CCC ansatz used for preparing state $\vert D^3_2\rangle$ is transformed into the CCC ansatz (followed by $X^{\otimes 3}$) for preparing state $\vert D^3_1\rangle$ based on the following circuit identities
\begin{equation}
\begin{split}
(X\otimes X){\rm CNOT}(X\otimes X) &= 
\begin{pmatrix}
0  &0   &1  &0 \\
0  &1   &0  &0 \\
1  &0   &0  &0 \\
0  &0   &0  &1
\end{pmatrix},\\
\label{eq:zero controlled Ry and NOT}
\end{split}
\end{equation}

\begin{equation}
\begin{split}
(X\otimes X)({\rm C-RY}(\theta))(X\otimes X) &= 
\begin{pmatrix}
\cos(\frac{\theta}{2})  &\sin(\frac{\theta}{2})  &0  &0 \\
-\sin(\frac{\theta}{2}) &\cos(\frac{\theta}{2})   &0  &0 \\
0               &0               &1  &0 \\
0               &0               &0  &1
\end{pmatrix}.
\label{eq:zero controlled Ry and NOT2}
\end{split}
\end{equation}
In other words, the $X$ gates in the first time step are transferred to the last time step. Consequently, the 3C block composed of zero-controlled CNOT and C-RY gates in Fig. \ref{eq:zero controlled Ry and NOT}(b) is
\begin{equation}
\begin{split}
\tilde{V}_i = 
\begin{pmatrix}
1 &0                 &0                &0 \\
0 &\cos(\theta_i/2)  &-\sin(\theta_i/2)    &0 \\
0 &\sin(\theta_i/2) &\cos(\theta_i/2)    &0 \\
0 &0                 &0                &1
\end{pmatrix}.
\label{eq:zero controlled 3C block V_i}
\end{split}
\end{equation}
As can be seen, $\tilde{V}_i$ is the adjoint of $V_i$, i.e. $\tilde{V}_i = V^{\dagger}_i$, see Eq. \eqref{eq:matrix representation of Vi}. Therefore, they are equivalent when their parameters are opposite. Finally, state $\vert D^3_1\rangle$ is converted to $\vert D^3_2\rangle$ by the $X^{\otimes 3}$ gates in the last time step.

In this paper, we assume that the operator $D^n_k=U_n((I\otimes X)^{\otimes k}\otimes I^{\otimes n-2k})$ represents the CCC ansatz for preparing $\vert D^n_k\rangle$, with $k\leq \frac{n}{2}$. Then, we have
\begin{widetext}
\begin{equation}
\begin{split}
X^{\otimes n}D^n_k &= X^{\otimes n}\overline{V}_{k-1}(\overline{V}_{k}\otimes \overline{V}_{k-2})(\overline{V}_{k+1}\otimes \overline{V}_{k-1}\otimes \overline{V}_{k-3})...(\overline{V}_{n-2}\otimes \overline{V}_{n-4} \otimes ... \otimes \overline{V}_{n-2k})((I\otimes X)^{\otimes k}\otimes I^{\otimes n-2k}) \\
&= \overline{V}^{\dagger}_{k-1}(\overline{V}^{\dagger}_{k}\otimes \overline{V}^{\dagger}_{k-2})(\overline{V}^{\dagger}_{k+1}\otimes \overline{V}^{\dagger}_{k-1}\otimes \overline{V}^{\dagger}_{k-3})...(\overline{V}^{\dagger}_{n-2}\otimes \overline{V}^{\dagger}_{n-4} \otimes ... \otimes \overline{V}^{\dagger}_{n-2k})((X\otimes I)^{\otimes k}\otimes X^{\otimes n-2k}) \\
&= D^n_{n-k},
\label{eq:U^n_k to U^n_{n-k}}
\end{split}
\end{equation}
\end{widetext}
where $V_i$ is abbreviated to $\overline{V}$, and the subscript $q$ in $\overline{V}_q$ represents the lower qubit which $\overline{V}_q$ operates on, refer to ref. \cite{wang2024vqe-po} for the detailed construction of $D^n_k$. In the first time step, replacing $(X\otimes I)^{\otimes k}\otimes X^{\otimes n-2k}$ by $(I\otimes X)^{\otimes k}\otimes X^{\otimes n-2k}$ does not affect the completeness of the output basis states. In summary, the CCC ansatz for preparing $\vert D^n_k\rangle$ can be transformed into that for preparing $\vert D^n_{n-k}\rangle$ and vice versa. Therefore, they have the same completeness.

Now, we proof the completeness and reachability of the folded staircase structure CCC ansatz $D^n_k$ for preparing $\vert D^n_k\rangle$ with $k\leq \frac{n}{2}$. Cases with $k>\frac{n}{2}$ can be directly proved based on $D^n_k=X^{\otimes n}D^n_{n-k}$. The procedure is summarized as

\begin{equation}
\begin{split}
\begin{pmatrix}
&D^2_1 &D^3_1 & & & \cdots & & &D^{n-1}_1 &D^n_1 \\
& & &D^4_2 &D^5_2 & & \cdots & &D^{n-1}_2 &D^n_2 \\
& & & &\ddots & & & & & \vdots \\
& & & & &D^{2k}_k &D^{2k+1}_k & \cdots &D^{n-1}_k &D^n_k \\
& & & & & &\ddots & & & \vdots \\
& & & & & & &D^{n-2}_{n/2-1} &D^{n-1}_{n/2-1} &D^n_{n/2-1} \\
& & & & & & & & &D^n_{n/2}
\end{pmatrix}.
\label{eq:proof process of completeness and reachability}
\end{split}
\end{equation}

\vspace{10pt}
\noindent \textbf{Case 1: ansatz \boldmath $D^n_1$}

\begin{figure*}[ht]
    \centering
    \includegraphics[width=0.75\textwidth]{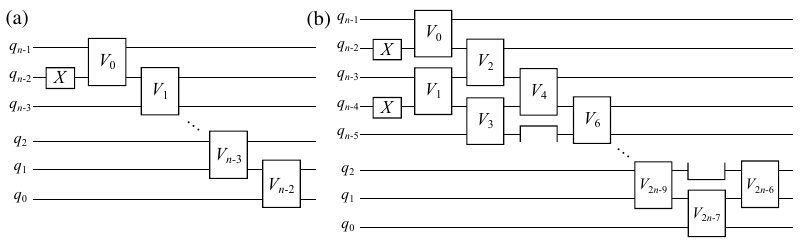}
    \caption{The folded staircase structure CCC ansatze for (a) $D^n_1$, (b) $D^n_2$.}
    \label{fig: Dn1n2 completeness and reachability}
\end{figure*}

(1) Based on Eq. \eqref{eq:matrix representation of Vi} and Fig. \ref{fig: Dn1n2 completeness and reachability}(a),
\begin{equation}
\begin{split}
D^2_1\vert 00\rangle = &V_0(I\otimes X)\vert 00\rangle = V_0\vert 01\rangle \\
= &\cos(\theta_0/2)\vert 01\rangle -\sin(\theta_0/2)\vert 10\rangle = \vert D^2_1\rangle.
\label{eq: D^2_1 completeness and reachability}
\end{split}
\end{equation}
This proves the completeness of $D^2_1$. $\vert 01\rangle$ is obtained with $100\%$ probability when $\theta_0 = 0$, while $\vert 10\rangle$ is obtained with $100\%$ probability when $\theta_0 = \pi$. This proves the reachability of $D^2_1$.

(2) For $D^3_1$, based on Eq. \eqref{eq:induction sum up} and Fig. \ref{fig: Dn1n2 completeness and reachability}(a),
\begin{equation}
\begin{split}
D^3_1\vert 0\rangle^{\otimes 3} &= (I\otimes V_1)(D^2_1\otimes I)\vert 0\rangle^{\otimes 3} = I\otimes V_1\vert D^2_1\rangle\vert 0\rangle \\
&=_c I\otimes V_1(\vert D^1_1\rangle\vert D^1_0\rangle + \vert D^1_0\rangle\vert D^1_1\rangle)\vert 0\rangle \\
&=_c \vert D^1_1\rangle\vert D^2_0\rangle + \vert D^1_0\rangle\vert D^2_1\rangle =_c \vert D^3_1\rangle.
\label{eq: D^3_1 completeness and reachability}
\end{split}
\end{equation}
This proves the completeness of $D^3_1$. State $\vert D^3_1\rangle$ can also be decomposed as $\vert D^3_1\rangle =_c \vert D^2_1\rangle\vert 0\rangle + \vert D^2_0\rangle\vert 1\rangle$. Therefore, $\vert D^2_1\rangle\vert 0\rangle$ can be obtained with $100\%$ probability when $\theta_1 = 0$, while $\vert D^2_0\rangle\vert 1\rangle$ can be obtained with $100\%$ probability when $\theta_0 = 0$ and $\theta_1 = \pi$. Additionally, each basis state in $\vert D^2_1\rangle$ can be obtained with $100\%$ probability under different values of $\theta_0$, the reachability of $D^3_1$ is proved.

(3) Assume $D^{n-1}_1$ is complete and reachable, then
\begin{equation}
\begin{split}
D^n_1\vert 0\rangle^{\otimes n} &= (I^{\otimes n-2}\otimes V_{n-2})(D^{n-1}_1\otimes I)\vert 0\rangle^{\otimes 2} \\
&= I^{\otimes n-2}\otimes V_{n-2}\vert D^{n-1}_1\rangle\vert 0\rangle \\
&=_c I^{\otimes n-2}\otimes V_{n-2}(\vert D^{n-2}_1\rangle\vert D^1_0\rangle + \vert D^{n-2}_0\rangle\vert D^1_1\rangle)\vert 0\rangle \\
&=_c \vert D^{n-2}_1\rangle\vert D^2_0\rangle + \vert D^{n-2}_0\rangle\vert D^2_1\rangle =_c \vert D^n_1\rangle.
\label{eq: D^n_1 completeness and reachability}
\end{split}
\end{equation}
This proves the completeness of $D^n_1$. Relation $\vert D^n_1\rangle =_c \vert D^{n-1}_1\rangle\vert 0\rangle + \vert D^{n-1}_0\rangle\vert 1\rangle$ indicates that $\vert D^{n-1}_1\rangle\vert 0\rangle$ can be obtained with $100\%$ probability when $\theta_{n-2} = 0$. And $\vert D^{n-1}_0\rangle\vert 1\rangle$ can be obtained with $100\%$ probability when $\theta_0 = 0$ and $\theta_i = \pi$ with $i\in[1,n-2]$. As $D^{n-1}_1$ is reachable, the reachability of $D^n_1$ is finally proved. Additionally, based on Eq. \eqref{eq:U^n_k to U^n_{n-k}}, $D^n_{n-1}$ is also complete and reachable.

\vspace{10pt}
\noindent \textbf{Case 2: ansatz \boldmath $D^n_2$}

(1) As depicted in Fig. \ref{fig: Dn1n2 completeness and reachability}(b), when $\theta_1 = 0$, operator $D^4_2$ reduces to $D^3_1\otimes D^1_1$. When $\theta_1 = \pi$, operator $D^4_2$ reduces to $D^3_2\otimes D^1_0$. As proven above, $D^3_1$ and $D^3_2$ are both complete and reachable. According to $\vert D^4_2\rangle =_c \vert D^3_1\rangle\vert D^1_1\rangle + \vert D^3_2\rangle\vert D^1_0\rangle$, $D^4_2$ is complete and reachable.

(2) When $\theta_3 = 0$, operator $D^5_2$ reduces to $D^4_2\otimes D^1_0$. When $\theta_1 = 0$ and $\theta_3 = \pi$, operator $D^5_2$ reduces to $D^4_1\otimes D^1_1$. Based on $\vert D^5_2\rangle =_c \vert D^4_2\rangle\vert D^1_0\rangle + \vert D^4_1\rangle\vert D^1_1\rangle$, the completeness and reachability of $D^5_2$ is proved.

(3) Assume $D^{n-1}_2$ is complete and reachable. When $\theta_{2n-7} = 0$, operator $D^n_2$ reduces to $D^{n-1}_2\otimes D^1_0$. When $\theta_1 = 0$ and $\theta_i = \pi$ with odd $i \in [3,2n-7]$, operator $D^n_2$ reduces to $D^{n-1}_1\otimes D^1_1$. Finally, the completeness and reachability of $D^n_2$ is proved. Additionally, based on Eq. \eqref{eq:U^n_k to U^n_{n-k}}, $D^n_{n-2}$ is also complete and reachable.

\begin{figure*}[ht]
    \centering
    \includegraphics[width=0.8\textwidth]{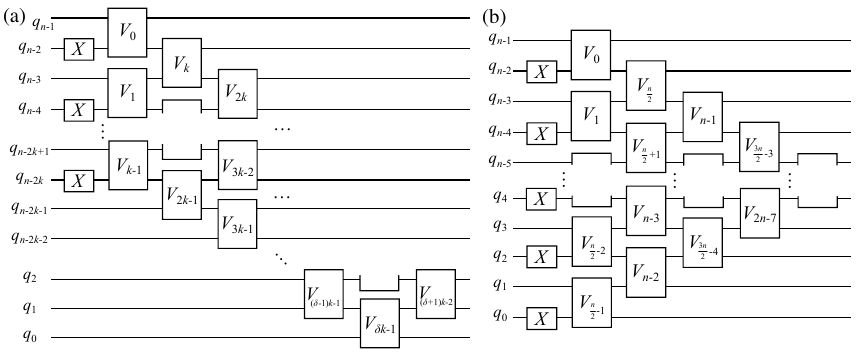}
    \caption{The folded staircase structure CCC ansatze for (a) $D^n_k$, (b) $D^n_{n/2}$. The parameter $\delta$ equals to $n-2k+1$.}
    \label{fig: Dnknn2 completeness and reachability}
\end{figure*}

\vspace{10pt}
\noindent \textbf{Case 3: ansatz \boldmath $D^n_k$ with $2\leq k\leq \frac{n}{2}-1$}

As we have already proved, operators on rows 1 and 2 of Eq. \eqref{eq:proof process of completeness and reachability} are complete and reachable. Therefore, we can assume the operators on row $k-1$ are complete and reachable. Based on $\vert D^{2k}_k\rangle =_c \vert D^{2k-1}_k\rangle\vert D^1_0\rangle + \vert D^{2k-1}_{k-1}\rangle\vert D^1_1\rangle$ and $D^{2k-1}_k = X^{\otimes 2k-1}D^{2k-1}_{k-1}$, the completeness and reachability of $D^{2k}_k$ is proved. Then, based on $\vert D^{2k+1}_k\rangle =_c \vert D^{2k}_k\rangle\vert D^1_0\rangle + \vert D^{2k}_{k-1}\rangle\vert D^1_1\rangle$, the completeness and reachability of $D^{2k+1}_k$ is proved. 

Now we can assume $D^{n-1}_k$ is complete and reachable. As depicted in Fig. \ref{fig: Dnknn2 completeness and reachability}(a), when $\theta_{\delta k-1} = 0$, operator $D^n_k$ reduces to $D^{n-1}_k\otimes D^1_0$. When $\theta_{k-1} = 0$ and $\theta_{ik-1} = \pi$ with $i \in [2,\delta]$, operator $D^n_k$ reduces to $D^{n-1}_{k-1}\otimes D^1_1$. Based on $\vert D^n_k\rangle =_c \vert D^{n-1}_k\rangle\vert D^1_0\rangle + \vert D^{n-1}_{k-1}\rangle\vert D^1_1\rangle$, the completeness and reachability of $D^n_k$ is proved. Based on Eq. \eqref{eq:U^n_k to U^n_{n-k}}, $D^n_{n-k}$ is complete and reachable.

\vspace{10pt}
\noindent \textbf{Case 4: ansatz \boldmath $D^n_{n/2}$}

In fact, ``\textbf{Case 3}'' with $k=\frac{n}{2}$ can prove the completeness and reachability of $D^n_{n/2}$. In this paper, $D^n_{n/2}$ is the main focus. Hence, its proof is listed separately. Assume the operators on row $\frac{n}{2}-1$ of Eq. \eqref{eq:proof process of completeness and reachability} are complete and reachable. As depicted in Fig. \ref{fig: Dnknn2 completeness and reachability}(b), when $\theta_{\frac{n}{2}-1} = 0$, operator $D^n_{n/2}$ reduces to $D^{n-1}_{n/2-1}\otimes D^1_1$. When $\theta_{\frac{n}{2}-1} = \pi$, operator $D^n_{n/2}$ reduces to $D^{n-1}_{n/2}\otimes D^1_0$. Based on $\vert D^n_{n/2}\rangle =_c \vert D^{n-1}_{n/2}\rangle\vert D^1_0\rangle + \vert D^{n-1}_{n/2-1}\rangle\vert D^1_1\rangle$ and $D^{n-1}_{n/2} = X^{\otimes n-1} D^{n-1}_{n/2-1}$, the completeness and reachability of $D^n_{n/2}$ is proved.

In conclusion, an arbitrary Dicke state $\vert D^n_k\rangle$, with $k\in [1,n-1]$, prepared by the folded staircase structure CCC ansatz \cite{wang2024vqe-po} is complete and reachable. In fact, there still exits redundancies in the folded staircase structure CCC ansatz. For instance, parameter configuration $\theta_0=0,\theta_1=\pi$ with arbitrary $\theta_2\in[0,\pi]$ can always obtain the basis state $\vert 0110\rangle$ from $\vert D^4_2\rangle$ with $100\%$ probability. Due to the more degrees of freedom of straight staircase structure CCC ansatze, its completeness and reachability is guaranteed without strain.

\section{\label{appen:principal component upon parameters}Relationship between parameters and the position of the principal component}
Here, we analyze the relationship between the parameters and the position of the principal component of the Dicke state, based on the propagation of ``$1$'' information in the ansatz. The analysis is specific to the folded staircase CCC ansatz illustrated in Fig. \ref{fig:figure prodecomp}(a).

For initial states $\vert 00\rangle$, $\vert 01\rangle$, $\vert 10\rangle$, and $\vert 11\rangle$, the states prepared by $V_i$ in Eq. \eqref{eq:matrix representation of Vi} with $i=0$ are
\begin{equation}
\begin{split}
V_0\vert 00\rangle &=\vert 00\rangle,\\
V_0\vert 01\rangle &=\cos(\theta_0/2)\vert 01\rangle -\sin(\theta_0/2)\vert 10\rangle, \\
V_0\vert 10\rangle &=\sin(\theta_0/2)\vert 01\rangle + \cos(\theta_0/2)\vert 10\rangle,\\
V_0\vert 11\rangle &=\vert 11\rangle.
\label{eq:V0 01 10}
\end{split}
\end{equation}
There are two cases: (1) $V_0$ acts as an identity operator when the initial state is $\vert 00\rangle$ or $\vert 11\rangle$; (2) $V_0$ prepares a superposition state composed of $\vert 01\rangle$ and $\vert 10\rangle$ when the initial state is $\vert 01\rangle$ or $\vert 10\rangle$. This means that, the position, rather than the count, of the $1$s in the input state may be altered by $V_i$. Additionally, the proportion of $\vert 01\rangle$ ($\vert 10\rangle$) in the output state, when the initial state is $\vert 01\rangle$ ($\vert 10\rangle$), depends on the cosine of the parameter $\frac{\theta_0}{2}$. In other words, the bit flipping from $\vert 01\rangle$ to $\vert 10\rangle$ ($\vert 10\rangle$ to $\vert 01\rangle$) occurs with probability $\sin^2(\frac{\theta_0}{2})$. In consequence, in a general CCC ansatz, the information represented by the $1$s in the initial state $\vert 0101...01\rangle$, can propagate through the circuit to generate a complete Dicke state. The amplitudes of its basis states are multiplications and summations of certain $\cos(\frac{\theta_i}{2})$ and $\sin(\frac{\theta_i}{2})$, where $\theta_i$ is the parameter of $V_i$.

Now, we use the preparation of Dicke state $\vert D^4_2\rangle$ as a warm-up example to illustrate the relationship between the parameters and the position of the principal component. The relationship depends on how the propagation of the ``$1$'' information occurs in the circuit. The circuit can be illustrated by the part constructed by the two $X$ gates and the three $2$-qubit unitaries $V_1,V_2,V_4$ (with parameters $\theta_0, \theta_1, \theta_2$ respectively) on qubits $q_3,q_2,q_1,q_0$, as shown in Fig. \ref{fig:figure prodecomp}(a). The propagation process is as follows:

(1) The initial state is prepared as $\vert 01\rangle\vert01\rangle$ by the two $X$ gates in the first time step.

(2) In the second time step, the operator $V_1 \otimes V_2$ (the first layer of $V_i$s) propagates the $1$s in the initial state $\vert 01\rangle\vert01\rangle$ to 
\begin{equation}
\begin{split}
V_1\otimes V_2\vert 01\rangle^{\otimes 2} =&(c_0\vert 01\rangle - s_0\vert 10\rangle)(c_1\vert 01\rangle - s_1\vert 10\rangle) \\
=& c_1c_0\vert 0101\rangle - c_1s_0\vert 0110\rangle \\
&- s_1c_0\vert 1001\rangle  + s_1s_0\vert 1010\rangle,
\label{eq:U4 propagation V1V2}
\end{split}
\end{equation}
where $\cos(\frac{\theta_i}{2})$ and $\sin(\frac{\theta_i}{2})$ are abbreviated to $c_i$ and $s_i$ respectively. As shown, the initial state $\vert 01\rangle\vert01\rangle$ is propagated to a linear combination of $\sum^1_{l,t=0} \vert l\rangle \otimes\vert t\rangle$ with $l,t\in\{01,10\}$. As $\vert l\rangle$ and $\vert t\rangle$ gradually change from $\vert 01\rangle$ to $\vert 10\rangle$, the basis state becomes larger and its probability amplitude contains more $s_i$. In consequence, as the identical parameter $\theta$ increases from $0$ to $\pi$, the principal component gradually migrates from $\vert 0101\rangle$ to $\vert 1010\rangle$.

(3) In the third time step, the operator $V_4$ is applied to the middle two qubits, $q_2,q_1$. In other words, the highest qubit and the lowest qubit remain unchanged. Eq. \eqref{eq:U4 propagation V1V2} evolves to
\begin{equation}
\begin{split}
V_4(V_1\otimes V_2)\vert 01\rangle^{\otimes 2} =& (s_2c_1c_0\vert 0011\rangle + c_2c_1c_0\vert 0101\rangle) \\
&- c_1s_0\vert 0110\rangle - s_1c_0\vert 1001\rangle\\
& + (c_2s_1s_0\vert 1010\rangle - s_2s_1s_0\vert 1100\rangle),
\label{eq:U4 propagation V4}
\end{split}
\end{equation}
where only the basis states containing $\vert 01\rangle$ or $\vert 10\rangle$ on the middle two qubits are propagated through $V_4$. The crucial point is that the propagation only occurs on the neighbours of the input basis state. For instance, the input $\vert 0101\rangle$ is propagated to $\vert 0011\rangle$, and $\vert 1010\rangle$ is propagated to $\vert 1100\rangle$, see Eq. \eqref{eq:U4 propagation V4}. The size of the neighbourhood is restricted by the position of the flipped bits in the basis state. Higher qubits own larger size of neighbourhood. For a Dicke state with a larger $n$, the layer by layer reduction of $V_i$s provides a local propagation capability of ``$1$'' information through the circuit. The local propagation contracts gradually towards the qubits in the middle. Therefore, the probability can only be redistributed in a local manner, which does not change the location of the principal component produced by the first layer of $V_i$s. This is summarized as the two phases for propagating the ``$1$'' information through the ansatz in the main text.

\begin{figure*}[ht]
    \centering
    \includegraphics[width=0.75\textwidth]{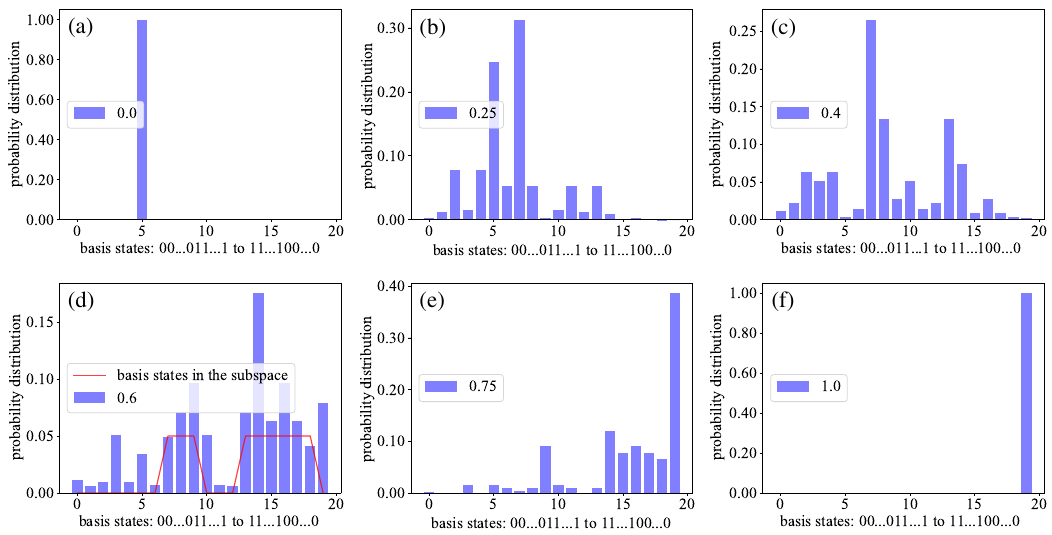}
    \caption{(a) – (f) The probability distributions prepared by the $\vert D^6_3\rangle$ ansatz with parameters set at $0$, $0.25\pi$, $0.4\pi$, $0.6\pi$, $0.75\pi$ and $\pi$. The principal component of $\vert D^6_3\rangle$ with parameter $0.6\pi$ covers the the basis states of sa$_2$, which is represented by the red line in (d).}
    \label{fig:figure paradistri}
\end{figure*}

From the perspective of the output probability distribution, illustrated in Fig. \ref{fig:figure paradistri}, as the value of the parameter increases within the interval [$0, \pi$], the principal component is gradually transferred from the initial basis state $\vert 0101...01\rangle$ to the final basis state $\vert 11...100...0\rangle$. Concurrently, the probability distribution gradually spreads across the entire space and then contracts. The inverse process occurs on the interval [$\pi, 2\pi$]. This shows the symmetry and periodicity of the probability distribution with respect to identical parameters. However, the probability distributions on the intervals [$0, \frac{\pi}{2}$] and [$\frac{\pi}{2}, \pi$] are not symmetrical to each other. On the interval [$0, \frac{\pi}{2}$], the initial basis state $\vert 0101...01\rangle$ is not the smallest basis state and the probability distribution experiences severe fluctuations. On the contrary, on the interval [$\frac{\pi}{2}, \pi$], the final basis state $\vert 11...100...0\rangle$ is the largest basis state, and the probability distribution is much smoother. This is consistent with the analyse result based on ratio $\delta$ and variance $\sigma$ in the main text. In consequence, we choose values from the interval [$\frac{\pi}{2}, \pi$) to initialize the parameters heuristically. The values around $\pi$ are excluded to avoid the identity matrix initialization of $U_n$, which produces bad convergences in the present scenario.

\section{\label{appen:complexity of RAEP}Complexity analysis of recursive ansatz equilibrium partition}
The Hamming distance between any two basis states $\vert x\rangle$ and $\vert y\rangle$ of a Dicke state $\vert D^n_k\rangle$ is an even number no less than 2, denoted as HD$(\vert x\rangle,\vert y\rangle)\geq 2$. Consequently, the entanglement entropy between $\vert x\rangle$ and $\vert y\rangle$ is not zero. For instance, in state $\vert \psi\rangle=a\vert 01\rangle + b\vert 10\rangle$ ($\vert \psi\rangle=a\vert 00\rangle + b\vert 11\rangle$), the entanglement entropy between $\vert 01\rangle$ and $\vert 10\rangle$ ($\vert 00\rangle$ and $\vert 11\rangle$) is
\begin{equation}
\begin{split}
S(\vert \psi\rangle\langle \psi\vert) =& -{\rm tr}(\vert \psi\rangle\langle \psi\vert {\rm log}\vert \psi\rangle\langle \psi\vert) \\
=&-\vert a\vert^2 {\rm log}\vert a\vert^2 - \vert b\vert^2{\rm log}\vert b\vert^2.  % \rm log变直体
\label{eq:entanglement entropy}
\end{split}
\end{equation} %%%%%%%%%%%%%
In fact, when $\vert a\vert^2 = 1/2$, $\vert \psi\rangle$ is a maximally entangled state. Therefore, any two basis states composing the $n$-qubit composite Dicke state are entangled. They cannot be reconfigured as a separable state in each $1$-qubit subsystem. This implies that a complete partition should produce $\binom{n}{k}$ sub-ansatze, each of which only prepares a single and unique basis state with Hamming weight $k$. 

Now assuming $n=2^m$. Based on the recursive equilibrium partition, the first partition produces two $2^{m-1}$-qubit fragments, the second partition produces four $2^{m-2}$-qubit fragments, $...$, the $m$th partition produces $n$ $1$-qubit fragments. Similarly, for $n' =2^{m+1}$, the number of partitions is $m+1$. For $n<n'' <n' $, after $m$ partitions, there remains at least one fragment (and at most $n-1$ fragments) composed of 2 qubits. Hence, one more partition is required. In summary, recursively partitioning $\lceil$log($n$)$\rceil$ times results in a complete partition.

A rigorous analysis of the partition complexity of the highly entangled Dicke state is a complex permutation and combination problem, which may be NP-hard. In this appendix, we provide a loose upper bound for the total number of sub-ansatze, suitable for scenarios with a small number of partitions:

(1) In the first partition, Dicke state $\vert D^n_{n/2}\rangle$ is partitioned into $n/2+1$ sub-ansatze, referring to Eq. \eqref{eq:Dicke linear combination}. The most complex sub-ansatz is $\bigotimes^2_{i = 1} \vert D^{n/2}_{n/4}\rangle$.

(2) In the second partition, the product Dicke state $\bigotimes^2_{i = 1} \vert D^{n/2}_{n/4}\rangle$ is partitioned into $(n/4+1)^2$ sub-ansatze. A loose upper bound of the total number of sub-ansatze for the $n/2+1$ sub-ansatze produced in the first partition is $(n/2+1)(n/4+1)^2$. At this stage, the most complex sub-ansatz is $\bigotimes^4_{i = 1} \vert D^{n/4}_{n/8}\rangle$.

(3) In the third partition, the product Dicke state $\bigotimes^4_{i = 1} \vert D^{n/4}_{n/8}\rangle$ is partitioned into $(n/8+1)^4$ sub-ansatze, resulting in a loose upper bound of $(n/2+1)(n/4+1)^2(n/8+1)^4$.

(4) In this nested manner, the loose upper bound for the $p$th partition accumulates to $O(\prod^p_{i=1} (\frac{n}{2^i})^{2^{i-1}})$, which can be reformulated as
\begin{equation}
\begin{split}
\prod^p_{i=1} (\frac{n}{2^i})^{2^{i-1}} =& (\frac{n}{2^1})^{2^0}(\frac{n}{2^2})^{2^1}...(\frac{n}{2^p})^{2^{p-1}} \\
=& \frac{n^{2^0+2^1+...+2^{p-1}}}{2^{1\cdot2^0+2\cdot2^1+3\cdot2^2+...+p\cdot2^{p-1}}} \\
=& \frac{n^{2^p-1}}{2^{(p-1)2^p+1}}.
\label{eq:loose upper bound reformulation}
\end{split}
\end{equation}
Clearly, the power of the complexity function increases exponentially with $p$ when $p\ll {\rm log}(n)$. For loglog($n$) partitions, the loose upper bound for the total number of sub-ansatze is $O(\frac{n^{{\rm log}(n)-1}}{2^{({\rm loglog}(n)-1){\rm log}(n)+1}})$. Furthermore, the convex interpolation idea offers a way to circumvent exhaustive execution of the exponential sub-ansatze. Here, we analyze the number of sub-ansatze required to evaluate the ground state of $\vert D^n_{n/2}\rangle$ using convex interpolation. We consider the outer $4$ sub-ansatze, sa$^1_1$, sa$^1_2$, sa$^1_{n/2-2}$, sa$^1_{n/2-1}$, and the most complex sub-ansatz at the middle, sa$^1_{n/4}$, to give an asymptotic upper bound for the complexity. Here, we assume the interpolations are accurate. The states prepared by the $5$ sub-ansatze can be expressed as
\begin{equation}
\begin{split}
\vert \rm sa^1_1\rangle &= \vert D^{n/2}_1\rangle \otimes \vert D^{n/2}_{n/2-1}\rangle, \\
\vert \rm sa^1_2\rangle &= \vert D^{n/2}_2\rangle \otimes \vert D^{n/2}_{n/2-2}\rangle, \\
\vert {\rm sa}^1_{n/4}\rangle &= \vert D^{n/2}_{n/4}\rangle \otimes \vert D^{n/2}_{n/4}\rangle, \\
\vert {\rm sa}^1_{n/2-2}\rangle &= \vert D^{n/2}_{n/2-2}\rangle \otimes \vert D^{n/2}_2\rangle, \\
\vert {\rm sa}^1_{n/2-1}\rangle &= \vert D^{n/2}_{n/2-1}\rangle \otimes \vert D^{n/2}_1\rangle,
\label{eq:sub-ansatze sa1 to n/2-1}
\end{split}
\end{equation}
where $n$ is assumed to be a very large number. It is evident that the partition complexity is dominated by sa$^1_{n/4}$. For instance, in the second partition, the number of sub-ansatze for sa$^1_{n/4}$ is $(n/4+1)^2$, while the total number of sub-ansatze for the $4$ outer sub-ansatze is merely a constant, $10$.

Here, we first assume that the minimum energies of the $4$ outer sub-ansatze, sa$^1_1$, sa$^1_2$, sa$^1_{n/2-2}$, and sa$^1_{n/2-1}$, are obtained by brute force. The evaluation procedure is divided into two steps:

\noindent \textbf{Step 1}: 

The minimum energies of the $4$ outer sub-ansatze are used to interpolate the convex curve. The location of the minimum point of the convex curve indicates the ground state location, which is assumed to be in the most complex sub-ansatz, sa$^1_{n/4}$, in the analysis.

\noindent \textbf{Step 2}: 

(1) When the ansatz is partitioned only once, the ground state can be found by directly executing sa$^1_{n/4}$. The total number of sub-ansatze used is $5$, refer to Eq. \eqref{eq:sub-ansatze sa1 to n/2-1}, rather than $n/2+1$. This represents the simplest case. 

(2) When the ansatz is partitioned twice, sa$^1_{n/4}$ would be partitioned into multiple sub-ansatze and cannot be executed directly. Here, we use the method illustrated in Appendix \ref{appen: Searching in hard-constraint} to evaluate the ground state location. As illustrated in Fig. \ref{fig:figure minimatrix}(a), the $4$ outer main diagonal sub-ansatze, sa$^2_{[0,0]}$, sa$^2_{[1,1]}$, sa$^2_{[n/4-1,n/4-1]}$, and sa$^2_{[n/4,n/4]}$, are used to interpolate the convex curve. However, at this point, the minimum point of the convex curve indicates a sub-ansatz 
sa$^2_{[i,i]}$ in the main diagonal, which may not be the ground state location. Then the cruciform-structure iteration, shown in Fig. \ref{fig:figure cruciform}, is performed to migrate to the ground state location. The first iteration requires the $4$ sub-ansatze adjacent to sa$^2_{[i,i]}$, while the later iterations only require the (at most) $3$ non-executed sub-ansatze adjacent to the current minimum location. To achieve a compromise between the number of sub-ansatze and the accuracy of the results, the number of iterations is set to log($n$). Therefore, the total number of sub-ansatze used is $5+(2^2-1)$log($n$)$ + 1$. Vast numerical simulations imply that log($n$) is a very loose upper bound for general cases.

(3) When the ansatz is partitioned three times, the $4$ outer main diagonal sub-ansatze of the worst case are sa$^3_{[0,0,0,0]}$, sa$^3_{[1,1,1,1]}$, sa$^3_{[\frac{n}{8}-1,\frac{n}{8}-1,\frac{n}{8}-1,\frac{n}{8}-1]}$, and sa$^3_{[\frac{n}{8},\frac{n}{8},\frac{n}{8},\frac{n}{8}]}$. The number of sub-ansatze used in this worst case is $5+(2^3-1)$log($n$)$ + 1$. Therefore, the total number of sub-ansatze used is nested as $(5+(2^2-1)$log($n$)$ + 1)(5+(2^3-1)$log($n$)$ + 1)$.

(4) Finally, partitioning $p$ times results in $\prod^p_{i=1} (6+(2^i-1){\rm log}(n))=O(2^{\frac{p+1}{2}}{\rm log}^p(n))$ sub-ansatze.

In consequence, when $p={\rm loglog}(n)$, the number of sub-ansatze used is upper bounded by $O(({\rm log}(n))^{{\rm loglog}(n)})$. Initially, we assumed the minimum energies of the $4$ outer sub-ansatze are obtained by brute force. For large problem sizes, the additional complexity for searching the minimum energies of the $4$ outer sub-ansatze based on the recursive partition method does not change the asymptotic complexity. In addition, the interpolated minimum point is commonly located not far from the rightmost outer sub-ansatz. Based on the above analysis, it is reasonable to believe that partitioning no more than loglog($n$) times under accurate enough interpolation has great potential to provide quantum speedups. In practice, more accurate and efficient interpolation methods are required, which we leave for future work.

\section{\label{appen:bounded CVaR expectation}Bounded CVaR$_\alpha$ expectation under parameter correlation}
In this appendix, we use the case of $\vert D^{16}_8\rangle$ as an example to illustrate the bounded phenomena of the CVaR expectation with respect to different numbers of correlated parameters. As we already know that confidence level $\alpha$ can bound a soft cap of $\alpha*100\%$ for the probability of obtaining the optimal solution. Equivalently, this means the CVaR expectation is bounded by a soft bottom. Consequently, to illustrate how parameter correlation impacts the soft bottom of the CVaR expectation, the effect of $\alpha$ should be taken into consideration. We perform the simulations with $4$ different confidence levels $\alpha = [0.01, 0.05, 0.1, 0.2]$ and $4$ different numbers of correlated parameters $\beta = [1, 10, 20, 40]$. The true value of the ground state energy is $-0.254$.

\begin{figure}[ht]
\centering
\includegraphics[width=0.48\textwidth]{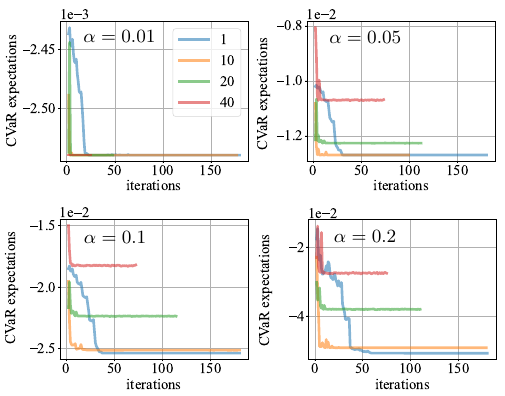}
    \caption{The CVaR expectation convergence curves with respect to different numbers of correlated parameters under different confidence levels $\alpha$.}
    \label{fig:figure boundcvar}
\end{figure}

On the one hand, as illustrated in Fig. \ref{fig:figure boundcvar}, for $\alpha = 0.01$, the bounded CVaR expectations of all $\beta$ are the same. At this point, the soft bottom is dominated by $\alpha$. As $\alpha$ increases, the CVaR expectations with larger $\beta$ converge much faster. They require significantly fewer iteration steps to achieve convergence precision. Therefore, initializing with smaller $\alpha$ and larger $\beta$ should be beneficial for the rapid descent of the cost expectation. On the other hand, as $\alpha$ increases, the bounded CVaR expectations of smaller $\beta$ descend lower. This corresponds to a spurious ground state that is a superposition state containing the true ground state with probably a larger probability. So migrating to larger $\alpha$ and smaller $\beta$ is preferred at the convergence stage.

\section{\label{appen: experimental configurations and performance, topology}Experimental configuration, qubit performance and topology}
The first portfolio optimization experiment is performed on real dataset (Tushare, https://tushare.pro/). The stocks in the asset pool are $600316$, $601808$, $688009$, $000959$, $600141$, $600010$, $000717$, $600500$, $002212$, $600776$, $600350$, $002048$ of China A-shares. The temporal range is selected between 2020/11/02 and 2020/12/31. The performance of an asset is represented by its closing price. The second portfolio optimization experiment is performed on random dataset (Qiskit finance) \cite{Qiskit}. The temporal range is selected between 2022/12/01 and 2022/12/30. The graph in the third experiment is generated by networkx \cite{networkx}. 
%\textcolor{red}{The number of measurement shots in all experiments is 2000.} 
As shown in Table \ref{tab:Experimental configurations}, the number of iterations of the COBYLA algorithm is set to $8$. That is to say, in the optimization process, the COBYLA algorithm is repeated $8$ times. The output parameters of the current COBYLA is applied as the initial parameters for the next iteration. The number of correlated parameters are gradually decreases to $1$. In other words, at the beginning of the optimization process, $8$ (or $6$) parameters (physical parameter) are correlated to behave as a single parameter (logical parameter). At this moment, the coarse cost function landscape is inundated with spurious local minima. As the number of correlated parameters decreases to $4$, $2$ (or $3$), the details of the cost function landscape become richer. Finally, the correlation vanishes when the number of correlated parameters decreases to $1$, refer to Appendix \ref{appen:bounded CVaR expectation} for the verification by numerical simulations. The ground state of the second experiment is located in sub-ansatz [$2,3$] of sub-ansatz $18$, refer to Appendix \ref{appen: Searching in hard-constraint} for the description of the presentation. 

\begin{table*}[ht] 
\setlength{\abovecaptionskip}{0.1cm}
\setlength{\belowcaptionskip}{0.1cm}
\caption{Experimental configurations.}
\label{tab:Experimental configurations}
\centering
\begin{threeparttable}
\scalebox{0.8}{
\renewcommand\arraystretch{1.5}{
\begin{tabular}{c| c |c c c}
\hline\hline
\multicolumn{2}{c}{Experiment} \vline & portfolio optimization $\vert D^{12}_6\rangle$  & portfolio optimization $\vert D^{40}_{20}\rangle$  & graph bisection $\vert D^{12}_6\rangle$\\
\multicolumn{2}{c}{Method} \vline & soft-constraint subspace  & hard-constraint subspace & soft-constraint subspace  \\
\multicolumn{2}{c}{Asset/graph seed} \vline & \textbackslash  & $1000$ & $1000$ \\
\multicolumn{2}{c}{Initial identical parameter} \vline & $0.65\pi$  & $0.8\pi$ & $0.75\pi$ \\
\multicolumn{2}{c}{Qubits} \vline & Q$14$-Q$17$\tnote{1}  & Q$14$-Q$10$ & Q$14$-Q$11$ \\
\multicolumn{2}{c}{\# of iterations} \vline & $8$  & $8$ & $8$  \\
\multirow{3}{*}{For each iteration}  & \# of correlated $\theta$s & [8,8,8,4,2,1,1,1] & [6,6,6,3,3,3,1,1] & [8,8,8,4,2,1,1,1] \\
                                    & \# of epochs & [15,12,10,15,19,31,31,31] & [24,24,24,38,38,38,39,39] & [15,12,10,12,18,30,30,30] \\
                                    & rho (*$\pi$) & \makecell{[0.15,0.136,0.124,0.113,\\0.102,0.07,0.07,0.07]} & \makecell{[0.15,0.15,0.15,0.15,\\0.15,0.1,0.1,0.1]} & \makecell{[0.15,0.136,0.124,0.113,\\0.102,0.1,0.1,0.1]} \\
\multicolumn{2}{c}{Ground state location} \vline & \textbackslash  & sa$^1_{18}$ - sa$^2_{[2,3]}$ & \textbackslash  \\ \hline\hline

\multicolumn{4}{l}{
\begin{minipage}{8cm}
 $^1$ depicted in Table \ref{tab:Performances of the 12 qubits} and Fig. \ref{fig:figure qtopology}.
\end{minipage}
}\\
\end{tabular}}}
\end{threeparttable}
\end{table*}

\begin{table}[ht]
\setlength{\abovecaptionskip}{0.1cm}
\setlength{\belowcaptionskip}{0.1cm}
\caption{The performance of the 12 qubits on 2024/02/20.}
\label{tab:Performances of the 12 qubits}
\centering
\begin{threeparttable}
\scalebox{0.75}{
\renewcommand\arraystretch{2}{
\begin{tabular}{c c c c c c c c c c c c c c c c c c c c c c c c c}  % 25列
\hline\hline
Qubit &\multicolumn{2}{c}{Q14} &\multicolumn{2}{c}{Q8} &\multicolumn{2}{c}{Q7} &\multicolumn{2}{c}{Q6} &\multicolumn{2}{c}{Q0} &\multicolumn{2}{c}{Q1} &\multicolumn{2}{c}{Q2} &\multicolumn{2}{c}{Q3} &\multicolumn{2}{c}{Q4} &\multicolumn{2}{c}{Q10} &\multicolumn{2}{c}{Q11} &\multicolumn{2}{c}{Q17} \\ \hline
$T_1$(\textmu s) &\multicolumn{2}{c}{10.43} &\multicolumn{2}{c}{11.07} &\multicolumn{2}{c}{8.03} &\multicolumn{2}{c}{7.18} &\multicolumn{2}{c}{11.30} &\multicolumn{2}{c}{18.64} &\multicolumn{2}{c}{10.49} &\multicolumn{2}{c}{20.68} &\multicolumn{2}{c}{10.81} &\multicolumn{2}{c}{35.47} &\multicolumn{2}{c}{17.27} &\multicolumn{2}{c}{11.08} \\
$T_2$(\textmu s) &\multicolumn{2}{c}{1.34} &\multicolumn{2}{c}{1.21} &\multicolumn{2}{c}{1.45} &\multicolumn{2}{c}{1.21} &\multicolumn{2}{c}{2.84} &\multicolumn{2}{c}{2.10} &\multicolumn{2}{c}{1.30} &\multicolumn{2}{c}{0.94} &\multicolumn{2}{c}{3.09} &\multicolumn{2}{c}{1.21} &\multicolumn{2}{c}{1.47} &\multicolumn{2}{c}{1.45} \\
$f_{00}$($\%$) &\multicolumn{2}{c}{93.32} &\multicolumn{2}{c}{88.96} &\multicolumn{2}{c}{95.62} &\multicolumn{2}{c}{95.42} &\multicolumn{2}{c}{95.36} &\multicolumn{2}{c}{92.34} &\multicolumn{2}{c}{93.56} &\multicolumn{2}{c}{89.64} &\multicolumn{2}{c}{91.80} &\multicolumn{2}{c}{95.28} &\multicolumn{2}{c}{94.70} &\multicolumn{2}{c}{92.32} \\
$f_{11}$($\%$) &\multicolumn{2}{c}{84.06} &\multicolumn{2}{c}{84.72} &\multicolumn{2}{c}{86.28} &\multicolumn{2}{c}{91.80} &\multicolumn{2}{c}{87.78} &\multicolumn{2}{c}{82.98} &\multicolumn{2}{c}{79.00} &\multicolumn{2}{c}{86.18} &\multicolumn{2}{c}{87.20} &\multicolumn{2}{c}{88.56} &\multicolumn{2}{c}{91.88} &\multicolumn{2}{c}{84.20} \\
$f_{sq}$($\%$)\tnote{1} &\multicolumn{2}{c}{99.85} &\multicolumn{2}{c}{99.76} &\multicolumn{2}{c}{99.80} &\multicolumn{2}{c}{99.81} &\multicolumn{2}{c}{99.65} &\multicolumn{2}{c}{99.82} &\multicolumn{2}{c}{99.76} &\multicolumn{2}{c}{99.58} &\multicolumn{2}{c}{99.88} &\multicolumn{2}{c}{99.78} &\multicolumn{2}{c}{99.77} &\multicolumn{2}{c}{99.76} \\
$f_{CZ}$($\%$) & \hspace{6pt} &\multicolumn{2}{r}{98.12} &\multicolumn{2}{r}{98.12} &\multicolumn{2}{r}{96.49} &\multicolumn{2}{r}{98.59} &\multicolumn{2}{r}{98.27} &\multicolumn{2}{r}{97.16} &\multicolumn{2}{r}{98.64} &\multicolumn{2}{r}{98.70} &\multicolumn{2}{r}{97.42} &\multicolumn{2}{r}{97.71} &\multicolumn{2}{r}{98.40} \\ \hline\hline

\multicolumn{10}{l}{
\begin{minipage}{4.5cm}
\vspace{-10pt}
 $^1$ \emph{sq}: single-qubit gate.
\end{minipage}
}\\
\end{tabular}}}
\end{threeparttable}
\end{table}

\begin{figure}[ht]
    \centering
    \includegraphics[width=0.4\textwidth]{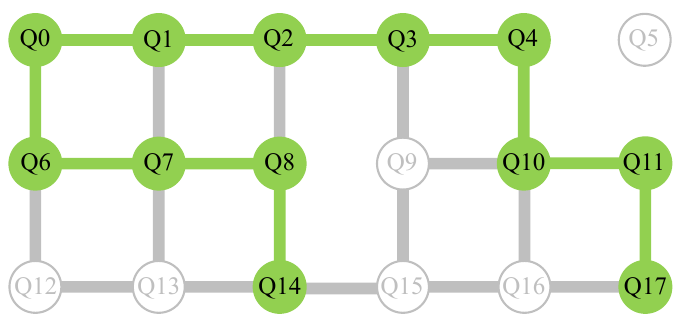}
    \caption{The topology of the 12 qubits.}
    \label{fig:figure qtopology}
\end{figure}

\section{\label{appen: Searching in hard-constraint}Hard-constraint subspace method example} %IV IV IV IV IV IV
In this Appendi, we provide the method (a process example) applied for evaluating the ground state location in the hard-constraint subspace scenario. The method is illustrated by an instance with Dicke state $\vert D^{40}_{20}\rangle$ and two recursions. Here, we assume the convex curve is accurately interpolated in the first recursion. 

\begin{figure*}[ht]
    \centering
    \includegraphics[width=0.88\textwidth]{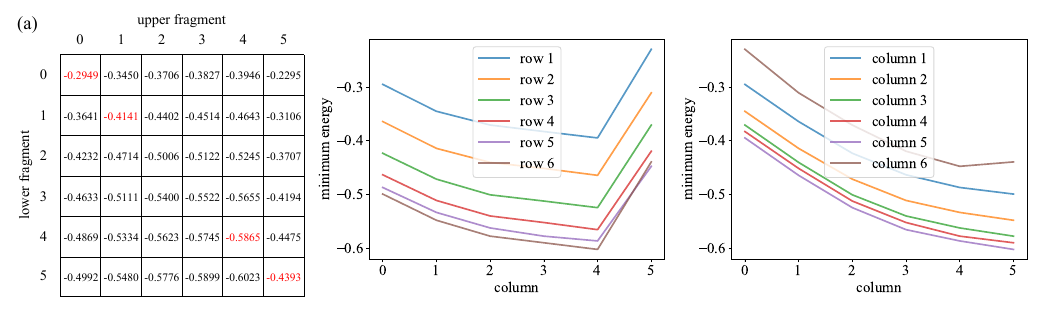}
    \caption{(a) The matrix used to store the minimum energies of the $36$ $40$-qubit sub-ansatze. The minimum energies are evaluated by the proposed methods. The seed for generating the asset pool is set to $1000$. And the risk level is set to $0.5$. (b) The interpolated convex curves for each row. The shape of the interpolated convex curves is almost the same. For all curves, the interpolated minima are located in column $4$. (c) The interpolated convex curves for each column. The location of the minimum of the curve of column $6$ is different from the others.}
    \label{fig:figure minimatrix}
\end{figure*}

(1) In the first recursion, the CCC ansatz is partitioned into $21$ sub-ansatze, i.e. sa$^1_0$, sa$^1_1$, $...$, sa$^1_{20}$, based on Eq. \eqref{eq:induction sum up reformulated}. For convenience, we use the superscript ``$i$'' to indicate the sub-ansatze obtained in the $i$th recursion. The minimum energies of the sub-ansatze sa$^1_1$, sa$^1_2$, sa$^1_{19}$, and sa$^1_{20}$, are used to interpolate the convex curve. Assuming the ground state is evaluated to be located in sa$^1_{15}$.  The corresponding product state is $\vert D^{20}_{15}\rangle\otimes\vert D^{20}_5\rangle$. The upper and lower fragments contain $15$ and $5$ $X$ gates respectively.

(2) In the second recursion, each fragment is partitioned into $6$ $20$-qubit sub-ansatze ($\vert D^{20}_{15}\rangle$ is converted to $\vert D^{20}_5\rangle$ by $X^{\otimes 20}$). In consequence, the production of the two fragments produces $36$ $40$-qubit sub-ansatze. Obviously, executing these sub-ansatze exhaustively results in poor performance. As illustrated in Fig. \ref{fig:figure minimatrix}, the distribution of the minimum energies of these sub-ansatze follows a similar pattern. The clustering property leads to the interpolation of similar convex curves for all rows (and columns). This inspires us to approximate the ground state location by curve interpolation twice. Firstly, interpolating a row to locate the best column, and then interpolating this column to find the ground state location. As shown in Fig. \ref{fig:figure minimatrix}(c), the distribution pattern could be violated. Additionally, the interpolated curve could also locate a wrong sub-ansatz. We need to check whether the minimum energies of the sub-ansatze adjacent to the located sub-ansatz are smaller.

\begin{figure}[ht]
    \centering
    \includegraphics[width=0.35\textwidth]{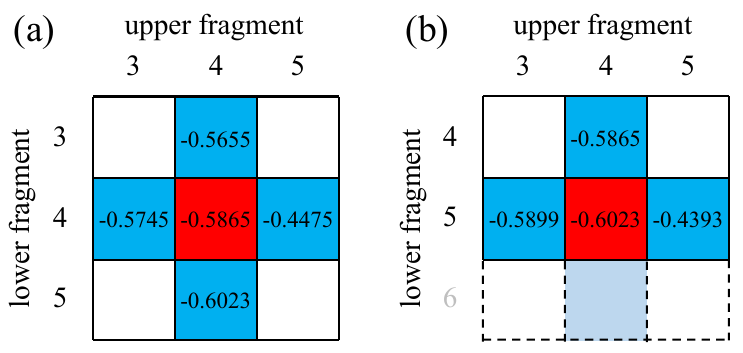}
    \caption{The greedy strategy used for migrating the minimum energy location. (a) The cruciform structure in the first iteration. The current minimum energy location migrates to $[4,5]$ as its minimum energy $-0.6023$ is the smallest one among the $5$ energies. (b) The reduced cruciform structure in the second iteration. The current minimum energy location remains at location $[4,5]$ as there are no smaller energies to migrate to. The iteration is terminated and the solution is the basis state with respect to the energy $-0.6023$.}
    \label{fig:figure cruciform}
\end{figure}

As can be seen, in a single row or column, the Hamming distance between the basis states in the contained sub-ansatze is only impact by the upper or the lower fragment. Smaller Hamming distance is more likely to deteriorate the interpolation accuracy. Hence, we choose to interpolate a single convex curve using the minimum energies of the $4$ outer sub-ansatze on the main diagonal of the matrix, i.e. elements $(0,0)$, $(1,1)$, $(4,4)$, $(5,5)$ as shown in Fig. \ref{fig:figure minimatrix}(a). They corresponds to sub-ansatze sa$^2_{[0,0]}$, sa$^2_{[1,1]}$, sa$^2_{[4,4]}$, and sa$^2_{[5,5]}$, respectively. Here, $i$ and $j$ in the subscript $[i,j]$ represent the $(i+1)$th and $(j+1)$th sub-sub-ansatz of $\vert D^{20}_{15}\rangle$ and $\vert D^{20}_{5}\rangle$, respectively. The Hamming distance between the basis states in these sub-ansatze is much larger as both the upper and lower fragments change. This should lead to a better interpolation and lower complexity. In this instance, the basis state corresponding to the interpolated minimum energy is located in sub-ansatz sa$^2_{[4,4]}$. As shown in Fig. \ref{fig:figure cruciform}, the red square indicates the \textbf{$\bm{c}$}urrent minimum energy location $[c_c, r_c]$ (The subscript ``$c$'' represents ``current''), and the blue squares indicate the adjacent sub-ansatze satisfying $|c_i- c_c|+|r_j- r_c|=1$. The red square and the adjacent blue squares construct a cruciform structure. As the cruciform structure migrates in the matrix, the ground state location can be approached iteratively. This is the greedy strategy used to indicate the minimum energy inside a sub-ansatz.

%\vspace{10pt}
\nocite{*}

\bibliography{apssamp}% Produces the bibliography via BibTeX.

\end{document}